\documentclass[12pt,a4paper,english]{article}
\usepackage{mypreamble}   

\usepackage{xr}
\setboolean{BLIND}{FALSE}   
\setboolean{APPON}{TRUE}   

\title{Diffusion index forecasts under weaker loadings: \\ PCA, ridge regression, and random projections}
\date{\today}

\ifthenelse{\boolean{BLIND}}{}
{
\author{Tom Boot\\
University of Groningen\\
\urlstyle{rm}\url{t.boot@rug.nl}\and Bart Keijsers\\
University of Amsterdam\\
\urlstyle{rm}\url{b.j.l.keijsers@uva.nl}}
}

\begin{document}
\maketitle

\spacingset{1.8} 

\begin{abstract}
    \noindent We study the accuracy of forecasts in the diffusion index forecast model with possibly weak loadings. The default option to construct forecasts is to estimate the factors through principal component analysis (PCA) on the available predictor matrix, and use the estimated factors to forecast the outcome variable. Alternatively, we can directly relate the outcome variable to the predictors through either ridge regression or random projections. We establish that forecasts based on PCA, ridge regression and random projections are consistent for the conditional mean under the same assumptions on the strength of the loadings. However, under weaker loadings the convergence rate is lower for ridge and random projections if the time dimension is small relative to the cross-section dimension. We assess the relevance of these findings in an empirical setting by comparing relative forecast accuracy for monthly macroeconomic and financial variables using different window sizes. The findings support the theoretical results, and at the same time show that regularization-based procedures may be more robust in settings not covered by the developed theory.\\

\end{abstract}

\newpage

\section{Introduction}
The diffusion index forecasting model developed by \citet{stock1998diffusion,stock2002forecasting} provides a parsimonious approach to use information from large macroeconomic data sets such as the monthly FRED-MD and quarterly FRED-QD \citep{mccracken2015fred,mccracken2020fred}. To establish consistency of diffusion index forecasts for the conditional mean, the classical assumption is that the factors are `pervasive' in the sense that these factors drive most variation in the available predictors. When there are 
$r$ factors, this pervasiveness is reflected in a sharp drop in the eigenvalues of the predictors' covariance matrix after the 
$r$th largest eigenvalue. In the absence of such a drop, PCA fails to consistently estimate the factors \citep{onatski2012asymptotics}, and hence, yields inconsistent forecasts.

Recently, \citet{bai2023approximate} provide an analysis of PCA-based factor estimation in the intermediate regime where the drop in the eigenvalues is `small'. In this paper, we utilize their results to analyze the consistency of diffusion index forecasts across three different estimation methods. First, we examine forecasts based on factors estimated by PCA. When the factor loadings are strong, the results naturally coincide with those in \citet{stock2002forecasting}. Under weaker loadings, we show that the convergence rate depends on the loadings' strength as well as on the relative growth of the cross-sectional dimension ($N$) and the time dimension ($T$). The convergence rate tends to drop when $N$ increases faster than $T$. We also show that a substantial improvement in the convergence rate can be obtained when the idiosyncratic errors in the factor model are temporally independent.

Next, we explore two alternative regularization-based forecast methods: ridge regression forecasts and random projection based forecasts. The former were analyzed under a factor model structure by \citet{de2008forecasting} as a form of Bayesian shrinkage. The latter can also be expected to work well when the data exhibits a factor structure \citep{boot2019forecasting}. We provide a unified analysis of both methods by deriving bounds on the implicit regularization of the forecast induced by the choice of the ridge penalty and the random projection dimension. Both methods effectively soft-threshold the eigenvalues of the predictor matrix, unlike PCA-based methods that apply hard-thresholding.

We find that appropriate selection of regularization parameters ensures the consistency of both regularization-based forecasts for the conditional mean under the same conditions as the PCA-based forecast. Under strong loadings, the results improve the convergence rates for ridge regression forecasts established by \citet{de2008forecasting}. We also show that the convergence rate can be negatively impacted by regularization bias under weaker loadings, particularly when the number of predictors grows faster than the sample size. In this case, the convergence rate can be slower relative to PCA. 

We provide two sets of simulations. The crucial difference between these simulations is the presence of serial correlation in the idiosyncratic errors. Without serial correlation, the simulations are in line with the asymptotic results even for small samples. If there is autocorrelation in the idiosyncratic errors, then for weak loadings and small sample sizes, the accuracy of PCA seems to falter, while that of the regularization methods stays remarkably stable. Part of the deterioration of PCA can be explained by the fact that in small samples the nominal number of factors may be lower than the optimal number of factors. When we increase the dimensions of the data, the accuracy of PCA again improves over ridge regression and random projections.

The simulations tie in with our empirical application on forecasting monthly and quarterly macroeconomic and financial series from the FRED-MD and FRED-QD \citep{mccracken2015fred,mccracken2020fred}. We estimate a diffusion index forecast model and manipulate the ratio of the sample size of the number of predictors in two ways while keeping the number of variables fixed: by choosing different estimation windows and by comparing across frequencies. 
In line with the developed theory, we find that as we decrease the moving window size, PCA forecasts become relatively more accurate compared to the regularization methods. At the same time, as in \citet{boot2019forecasting}, and our simulations results for small samples with serial correlation in the idiosyncratic errors, random projections and ridge frequently outperform the PCA-based forecast.
PCA yields better forecasts in about 10 percent of the cases when $T>N$, while PCA-based forecasts are more accurate than ridge and random projections for close to 40 percent of the variables when $T=\tfrac{1}{3}N$. In addition, in the monthly dataset, PCA-based forecasts are more accurate in 7 to 13 percent of the cases, but this increases to 19 percent when using quarterly data.

\paragraph{Related literature}
The possible presence of weak loadings, for example in macroeconomic and financial applications, has triggered an active literature. \citet{onatski2012asymptotics} considers an asymptotic regime where the eigenvalues are not well-separated, as in the spiked covariance matrix model of \citet{johnstone2009consistency} and \citet{paul2007asymptotics}. There, the ratio of the leading eigenvalues of the covariance matrix of the predictors with the smaller eigenvalues is assumed to be bounded. This constitutes an even weaker setting relative to the one considered in \citet{bai2023approximate} that we follow in this paper. 

Weak loadings as defined in the current paper can arise through a sparse structure of the loading matrix. Estimation and inference in this setting is considered in \citet{uematsu2022estimation} and \citet{uematsu2022inference}. \citet{fan2020learning} consider a setting with weak loadings that are not necessarily sparse. The approach is somewhat similar to the random projection forecast as they use a cross-sectional projection to estimate the factors. The results presented there require the weights of the projections to be bounded (Assumption 2.1.i in \citet{fan2020learning}). Also, instead of using a single projection, we consider averaging over multiple random projections to eliminate additional variance from the fact that we project onto a random subspace. \citet{karabiyik2020forecasting} consider simple averages of predictors within a pre-specified group to recover group-specific factors. In the context of the FRED-MD, the groups would be the predefined variable groups. The random projection forecast we consider, takes a randomly weighted average across the entire predictor set. 

Initial convergence rates for ridge regression in factor models as in \citet{de2008forecasting} and \citet{carrasco2016sample} suggested that ridge forecasts may converge at a slower rate compared to PCA-based forecasts. \citet{de2024asymptotic} show the asymptotic equivalence of PCA-based forecast and ridge regression in settings with possible weak factors. However, this requires additional assumptions that do not fit the factor model of \citet{bai2023approximate} that we consider in the present paper. We find that the equivalence between ridge forecasts and PCA-based forecasts holds under serially uncorrelated errors and a cross-section dimension not much larger than the time dimension. However, in other cases the regularization-based methods attain a slower convergence rate.

In a variant of the diffusion index forecast model where the idiosyncratic noise in the factor equation enters the outcome equation through a random effects specification, \citet{he2024ridge} shows the optimality of ridge regression. Importantly, the paper shows the validity of a bias-corrected cross-validation procedure to select the optimal penalty in practice. Our theoretical results determine the optimal penalty parameter up-to-scale, and in practice we set it based on historical forecast performance.

Whereas ridge regression has a long history in forecasting applications, the use of random projections is relatively recent. Random projections are generally motivated from the Johnson-Lindenstrauss lemma \citep{johnson1984extensions}, which can be used to show that they maintain a substantial fraction of the information, see for instance \citet{achlioptas2003database}. Random projections can be applied to discrete choice models \citep{chiong2016random}, to forecast product sales \citep{schneider2016forecasting}, to reduce the dimension of vector autoregressions \citep{koop2016bayesian}, to macroeconomic forecasting \citep{boot2019forecasting}, and  in high-dimensional hypothesis testing \citep{liu2023random}. 
The equivalence to ridge regression, in terms of the convergence rate, appears new.

\paragraph{Notation}  Denote $\delta_{NT} = \min(\sqrt{N},\sqrt{T})$ and the $j$th eigenvalue of a matrix $\bs A'\bs A$ as $\mu_{j}(\bs A'\bs A)$. $\mathcal{O}(T)$ denotes the space of orthogonal $T\times T$ matrices. Let, $||\bs A||^2=\text{tr}(\bs A'\bs A)$ denote the Frobenius norm. The vector $\bs b_{t}$ is the standard basis vector with a 1 in position $t$. Throughout, $M$ denotes a generic positive constant independent of any of the involved dimensions. For random variables $X$ and $Y$, we write $X\overset{(d)}{=}Y$ if $X$ and $Y$ have the same distribution. We write $T\asymp N$ if both $T/N=O(1)$ and $N/T = O(1)$.
\section{Model and forecasts}\label{sec:setup}
We consider the diffusion index forecast model by \citet{stock2002forecasting,stock2002macroeconomic} where a small number of factors  $\bs f_{t}\in \mathbb{R}^{r}$ drives the variable of interest, and the same factors underlie a  high-dimensional set of predictors $\bs x_{t}\in \mathbb{R}^{N}$,
\begin{eqnarray}
        \bs x_{t}& = \bs \Lambda\bs f_{t} + \bs e_{t},\label{eq:factorequation}\\
        y_{t+h}& = \bs f_{t}'\bs\gamma + \varepsilon_{t+h}.\label{eq:forecastequation}
\end{eqnarray}
Let $t=0,\ldots,T$, $\bs X=(\bs x_{0},\ldots,\bs x_{T-h})'$, $\bs F=(\bs f_{0},\ldots,\bs f_{T-h})'$, $\bs \Lambda = (\bs \lambda_{1}, \ldots,\bs\lambda_{N})'$, and $\bs y = (y_{h}, \ldots, y_{T})'$. One can add lagged outcome variables to \eqref{eq:forecastequation}, but we omit those from the analysis to keep the discussion focused on the effects of a weak factor structure. Throughout, we write $\bar{y}_{t+h}$ for the conditional mean of \eqref{eq:forecastequation}, i.e. $\bar{y}_{t+h} = \bs f_{t}'\bs\gamma$.

We follow \citet{bai2023approximate} in defining the rescaled matrix $\bs Z = \bs{X}/\sqrt{NT}$ that has the singular value decomposition
\begin{equation}\label{eq:defZ}
\bs{Z} = \bs U\bs D\bs V' = \bs U_{r}\bs D_{r}\bs V_{r}' +  \bs U_{-r}\bs D_{-r}\bs V_{-r}'.
\end{equation}
Here $\bs{U}\in \mathcal{O}(T)$, $\bs{D}$ is $T\times N$ with the $\delta_{NT}^2$ nonzero singular values of $\bs Z$ on the diagonal of its upper left $\delta_{NT}^2\times \delta_{NT}^2$ submatrix, and $\bs{V}\in\mathcal{O}(N)$. Moreover, $\bs D_{r}\in\mathbb{R}^{r\times r}$ is a diagonal matrix with the $r$ largest singular values on its diagonal, the columns of $\bs U_{r}\in\mathbb{R}^{T\times r}$ and $\bs V_{r}\in\mathbb{R}^{N\times r}$ contain the corresponding left- and right-singular vectors. The matrices $\bs U_{-r}$ and $\bs V_{-r}$ contain the remaining left- and right-singular vectors.

The standard assumption on the loadings accompanying \eqref{eq:factorequation} is that ${\bs \Lambda'\bs\Lambda/N\rightarrow_{p}\bs\Sigma_{\Lambda}}$ where $\bs\Sigma_{\Lambda}$ is a positive definite matrix. As in \cite{bai2023approximate}, this paper considers the case where $\bs \Lambda'\bs\Lambda/N^{\alpha}\rightarrow_{p}\bs\Sigma_{\Lambda}$ for some $\alpha\in (0,1]$. When $\alpha=1$, we say that the loadings are `strong', while when $\alpha<1$, we refer to the loadings as `weak'.

\subsection{PCA-based forecasts}
PCA-based forecasts rely on the following estimates of the factors and loadings,
\begin{equation}
(\tilde{\bs F},\tilde{\bs\Lambda}) = (\sqrt{T}\bs U_{r},\sqrt{N}\bs V_{r}\bs D_{r}).
\end{equation}
Denote by $\hat{\bs\gamma}$ the least squares estimator of $\bs\gamma$ in \eqref{eq:forecastequation} with $\bs f_{t}$ replaced by $\tilde{\bs f}_{t}=\tilde{\bs F}'\bs b_{t}$. The resulting PCA-based forecast is given by,
\begin{equation}\label{eq:pcaforecast}
\hat{y}_{T+h|T}^{pca} = \tilde{\bs f}_{T}'\hat{\bs\gamma},
\end{equation}
where
\begin{equation}\label{fac:est}
    \tilde{\bs f}_{T} = \frac{1}{\sqrt{N}}\bs D_{NT,r}^{-1}\bs V_{NT,r}'\bs x_{T}=O_{p}(1)\frac{1}{N^{\alpha}}\tilde{\bs\Lambda}'\bs\Lambda\bs f_{T} + O_{p}(1)\frac{1}{N^{\alpha}}\tilde{\bs\Lambda}'\bs e_{T}.
\end{equation}

\citet{stock2002forecasting} Theorem 2 shows that $\tilde{\bs f}_{T}'\hat{\bs\gamma}-\bs f_{T}'\bs\gamma\rightarrow_{p}0$ and the limiting distribution was established in \citet{bai2006confidence}. Under an approximate factor model with strong loadings, the convergence rate is $\delta_{NT}$.

\paragraph{Remark} There is a slight subtlety in the construction of the forecasts. Suppose that $\bs e_{t}$ is independent across $t$. In that case, in the final term of \eqref{fac:est}, $\bs e_{T}$ is independent of the error in estimating the loadings. This independence can be used to derive an improved convergence rate. An alternative way to construct the PCA-based forecast is to perform an eigendecomposition on the matrix formed by $(\bs x_{0}',\ldots,\bs x_{T}')$. However, in \eqref{fac:est}, this would lead to dependence between $\bs e_{T}$ and $\tilde{\bs\Lambda}$. With currently available results, this yields a slower convergence rate.

\subsection{Regularization-based forecasts} When the number of predictors is large, we can regularize the (potentially ill-defined) inverse of $\bs X'\bs X$ by adding a ridge penalty \citep{hoerl1970ridge}. This leads to the following forecast.
\begin{equation}\label{eq:ridgeforecast}
\hat{y}_{T+h|T}^{ridge} = \bs x_{t}'\left(\bs X'\bs X + \frac{NT}{k} \bs I_{N}\right)^{-1}\bs X'\bs y.
\end{equation}
This forecast is identical to a Bayesian forecast with a $N(\bs 0,\frac{k}{NT} \bs I_{N})$ prior on the regression coefficients. \citet{de2008forecasting} analyze the consistency of this forecast for the conditional mean $\bar{y}_{t+h}$ and find that it may converge at a slower rate compared to the PCA-based forecast \eqref{eq:pcaforecast}, even when loadings are strong. We will show in \cref{sec:theory} that the convergence rate of \eqref{eq:ridgeforecast} actually coincides with that of \eqref{eq:pcaforecast} under strong loadings.

\citet{boot2019forecasting} consider the following alternative forecasting model that is based on random projections of the available variables,
\begin{equation}\label{eq:modelrproj}
    y_{t+h,R} = \bs x_{t}'\bs R\bs\theta_{R} + \xi_{t+h},
\end{equation}
where $\bs R\in\mathbb{R}^{N\times k}$ and the element of $\bs R$ are independent standard normal random variables. We emphasize that \eqref{eq:modelrproj} is interpreted as a model that defines a forecast. The actual data generating process is always taken to be \eqref{eq:factorequation} and \eqref{eq:forecastequation}.

Denote by $\hat{\bs\theta}_{R}$ the least squares estimator of $\bs\theta_{R}$ in \eqref{eq:modelrproj} and denote the associated forecast as $\hat{y}_{T+h|T,R}^{rp}=\bs x_{T}'\bs R\hat{\bs\theta}_{R}$. The random projection introduces additional noise in the forecast, which can be reduced by averaging over multiple draws of the random matrix $\bs R$. The random projection forecast then approximates,
\begin{equation}\label{eq:rpforecast}
    \hat{y}_{T+h|T}^{rp} = \bs x_{T}'\mathbb{E}_{R}[\bs R\hat{\bs\theta}_{R}]=\bs x_{T}'\mathbb{E}_{R}[\bs R(\bs R'\bs X'\bs X\bs R)^{-1}\bs R']\bs X'\bs y.
\end{equation}
\citet{boot2019forecasting} show that if we take $O(N\log N)$ draws of the projection matrix $\bs R$, the mean squared forecast error approaches that of $\hat{y}_{T+h|T}^{rp}$.

It is not a coincidence that the regularization in both \eqref{eq:ridgeforecast} and \eqref{eq:rpforecast} is determined by the parameter $k$, as it is scaled such that the same $k$ (up to a constant scaling factor) determines the convergence rate to the conditional mean. In the subsequent sections, we use the notation $\hat{y}^{reg}_{T+h|T}$ to refer to both the ridge regression forecast and the random projection forecast.

\subsection{Regularizing the empirical eigenvalues}
For all three forecasts, the leading term in the expansion of the forecasts that captures the conditional mean $\bs f_{T}'\bs\gamma$ can be written as $
\bs f_{T}'(\bs F'\bs F)^{-1}\bs F'\bs U\bs \Delta\bs U'\bs F\bs\gamma.$
Suppose that we know that the true number of factors is $r$, and the regularization parameter $k$ is selected as we derive in \cref{sec:theory}. Then, $\bs \Delta$ is a diagonal matrix that satisfies the following.
\begin{center}
\begin{tabular}{lcc}
$\Delta_{ii}$& PCA & Ridge and RP\\
\midrule
$i=1,\ldots,r$& 1 & $1-O_{p}\left(\frac{N}{N^{\alpha}}k^{-1}\right)$ \\
$i = r+1,\ldots,\delta_{NT}^2$& 0& $O_{p}\left(k\delta_{NT}^{-2}\right)$
\end{tabular}
\end{center}
This clarifies that PCA is hard-thresholding the eigenvalues, while ridge regression and random projections are soft-thresholding the eigenvalues. However, in the asymptotic limit where $(N,T)\rightarrow\infty$, ridge regression and random projection may converge to PCA under a suitably chosen of $k$. To establish this formally requires a more detailed analysis of other terms appearing in the expansion of the forecasts. We also see that for weaker loadings, the regularization parameter $k$ has to be set larger to capture the signal in the first $r$ factors. However, at the same time this leads to a slower decline to zero for the remaining eigenvalues as this part is independent of the loading strength $\alpha$. This will eventually lower the convergence rate under weak loadings.

\section{Theory}\label{sec:theory}
We have the following assumptions.

\begin{assumption}\label{assA1}
    \begin{enumerate}[label=(\roman*)]
           \item The errors in \eqref{eq:factorequation} can be written as $\bs e = \bs R_{T}^{1/2}\bs a\bs G_{N}^{1/2},$ with $\bs R_{T}\in \mathbb{R}^{T\times T}$ and $\bs G_{N}\in \mathbb{R}^{N\times N}$ deterministic matrices,
        \item Conditional on $\bs\Lambda$ and $\bs F$, $a_{it}$ is i.i.d.\ over $i$ and $t$, with $\mathbb{E}[a_{it}|\bs\Lambda,\bs F]=0$, $\mathbb{E}[a_{it}^4|\bs\Lambda,\bs F]\leq M$,
        \item $\mu_{T}(\bs R_{T})\geq C$, $ \mu_{N}(\bs G_{N})\geq C$,
        \item\label{ass:gersh} $\max_{t}\sum_{s=1}^{T}|\bs R_{T,ts}|\leq C$,  $\max_{i}\sum_{j=1}^{N}|\bs G_{N,ij}|\leq C$.
    \end{enumerate}
\end{assumption}
This assumption is closely related to the assumptions used in \citet{onatski2010determining} and \citet{ahn2013eigenvalue}. Part \cref{ass:gersh} slightly strengthens the usual assumption on the maximum eigenvalue of $\bs R_{T}$ and $\bs G_{N}$, so we can easily establish that \cref{assA1} implies Assumption A1 and A3 in \citet{bai2023approximate}.

\begin{corollary}\label{corr1} Let $\bs e_{t}$ be defined as before and denote by $\bs e_{(i)}$ the $i$th column of $\bs e$. Under \cref{assA1} we have the following. 
    \begin{enumerate}[label=(\roman*)]
        \item For all $t$, $\frac{1}{N\sqrt{T}}\|\bs e_{t}'\bs e'\| = O_{p}(\delta_{NT}^{-1})$ and for all $i$, $\frac{1}{T\sqrt{N}}\|\bs e_{(i)}'\bs e\| =O_{p}(\delta_{NT}^{-1})\|$,
        \item For each $t$, $\mathbb{E}\|N^{-\alpha/2}\sum_{i=1}^{N}\bs\lambda_{i}e_{it}\|^2\leq C$ and $\frac{1}{NT}\bs e_{t}'\bs e'\bs F = O_{p}(\delta_{NT}^{-2})$,
        \item For each $i$, $\mathbb{E}\|T^{-1/2}\sum_{t=1}^{T}\bs f_{t}e_{it}\|^2\leq M$ and $\frac{1}{N^{\alpha}T}\bs e_{(i)}'\bs e\bs\Lambda = O_{p}\left(\frac{1}{N^{\alpha}}\right)+O_{p}\left(\frac{1}{\sqrt{N^{\alpha}T}}\right)$,
        \item $\bs\Lambda'\bs e'\bs F=\sum_{t=1}^{T}\sum_{i=1}^{N}\bs\lambda_{i}\bs{f}_{t}'e_{it}=O_{p}(\sqrt{N^{\alpha}T})$.
    \end{enumerate}
\end{corollary}
We continue with Assumption~A2 from \citet{bai2023approximate}.
\begin{assumption}\label{assA2}
\begin{enumerate}[label=(\roman*)]
\item $\mathbb{E}[\|\bs f_{t}\|^4]\leq M$, $\plim_{T\to\infty} \frac{\bs F'\bs F}{T}=\bs\Sigma_{F}\succ 0$,
\item $\|\bs\lambda_{i}\|\leq M$,  $\lim_{N\to\infty}\frac{\bs\Lambda'\bs\Lambda}{N^{\alpha}}=\bs\Sigma_{\Lambda}\succ 0$ for some $\alpha>0$ with $\alpha\in (0,1]$,
\item The eigenvalues of $\bs\Sigma_{\Lambda}\bs\Sigma_{F}$ are distinct.
\end{enumerate}
\end{assumption}
We also repeat Assumption A4 from \citet{bai2023approximate} that limits the expansion rate of the cross-section dimension relative to the time dimension.
\begin{assumption}\label{assA4}
    As $(N,T)\rightarrow\infty$, $\frac{N}{N^{\alpha}}\frac{1}{T}\rightarrow 0$, for $\alpha$ as before.
\end{assumption}

To derive results for the diffusion index model, we need to make assumptions on the regression errors $\varepsilon_{t}$, which are provided by the following assumption.
\begin{assumption}\label{ass:varepsilon} The regression errors $\bs \varepsilon$ satisfies $\mathbb{E}[\bs\varepsilon|\bs F,\bs\Lambda,\bs e]=\bs 0$ and  $\mathbb{E}[\bs\varepsilon\bs\varepsilon'|\bs F,\bs\Lambda,\bs e] = \bs\Sigma_{\varepsilon}$ with $\|\bs\Sigma_{\varepsilon}\|\leq M<\infty$ $a.s.$
\end{assumption}
Finally, several results that we derive can be improved under the following assumption that strengthens \cref{assA1} part (i) to ensure that the noise is independent over time.
\begin{assumption}\label{assA5}
    In \cref{assA1} part (i), set $\bs R_{T}=\bs I$.
\end{assumption}

\subsection{Convergence rates for PCA-based forecasts}
Under strong factors, \citet{stock2002forecasting} show consistency of the PCA forecast \eqref{eq:pcaforecast} for the conditional mean $\bar{y}_{T+h|T}=\bs f_{T}'\bs\gamma$ when $(N,T)\rightarrow\infty$. \citet{bai2006confidence} further show that $\hat{y}_{T+h|T}^{pca}-\bar{y}_{T+h|T} = O_{p}(\delta_{NT}^{-1})$.

Under weaker loadings, we have the following result for the PCA forecast.
\begin{theorem}\label{thm:0} Under \crefrange{assA1}{ass:varepsilon}, the PCA-based forecast \eqref{eq:pcaforecast} satisfies
\begin{equation}\label{eq:pca_gen}
\begin{split}
\hat{y}_{T+h|T}^{pca} - \bar{y}_{T+h|T} &=O_{p}\left(\frac{1}{N^{\alpha/2}}\right)+  O_{p}\left(\frac{1}{T^{1/2}}\right)+ O_{p}\left(\frac{N^{3/2}}{N^{3\alpha/2}T}\right)\\
&\quad +O_{p}\left(\frac{N^{1/2}}{N^{3\alpha/2}}\right) +  O_{p}\left(\frac{N^2}{N^{2\alpha}T^{3/2}}\right)+O_{p}\left(\frac{N}{N^{2\alpha}T^{1/2}}\right).
\end{split}
\end{equation}
\end{theorem}
The first two terms on the right-hand side reduce to the known results when $\alpha=1$. However, a number of additional terms appear. To interpret these, we analyze three cases of interest. In the first, the loadings are strong, so $\alpha=1$. In the second, the loadings are weaker $\alpha<1$, and $N\asymp T$. In the third setting, again $\alpha<1$, but now the time dimension is of lower order relative to the cross-section dimension in the sense that $T = O(N^{\gamma})$ with $\gamma<1$. Specialized to these cases, \cref{thm:0} yields the following.
\begin{corollary}\label{corr2}
    Case 1. Strong loadings: $\alpha=1$.
    \begin{equation}
         \hat{y}_{T+h|T}^{pca}- \bar{y}_{T+h|T} = O_{p}(\delta_{NT}^{-1}).
       \end{equation}
    Case 2. Weaker loadings: $\alpha<1$. $N\asymp T$.
\begin{equation}
 \hat{y}_{T+h|T}^{pca} - \bar{y}_{T+h|T} = O_{p}\left(\frac{1}{N^{\alpha/2}}\right) + O_{p}\left(\frac{1}{N^{(3\alpha-1)/2}}\right) .
\end{equation}
Case 3. Weaker loadings: $\alpha<1$. $T=O(N^{\gamma})$, $0<\gamma<1$.
\begin{equation}
    \begin{split}
        \hat{y}_{T+h|T}^{pca} - \bar{y}_{T+h|T}&= O_{p}\left(\frac{1}{N^{\alpha/2}}\right) + O_{p}\left(\frac{1}{N^{\gamma/2}}\right) +O_{p}\left(\frac{1}{N^{(3\alpha-3+2\gamma)/2}}\right).
    \end{split}
\end{equation}
\end{corollary}
Case 1 shows that under strong loadings, we obtain the known result on the convergence rate of PCA-based forecast. Case 2 shows that under weaker loadings, an additional term enters that requires that $\alpha>1/3$ for the forecast to remain consistent for the conditional mean. This result is not surprising, as \citet{bai2023approximate} already show that the same condition is required for (a rotation of) the factors $\bs F$ to be consistently estimated. We will later show that this restriction can be lifted under \cref{assA5}. Under Case 3, we see an interaction between the loading strength parameter $\alpha$ and the parameter $\gamma$ that governs how much smaller $T$ is relative to $N$. For consistency of the forecasts for the conditional mean, a sufficient condition is now that $\alpha>1-\frac{2}{3}\gamma$. This indicates that as the cross-sectional dimension increases relative to the number of time periods, we require stronger loadings for the forecast to be consistent.

\subsection{Convergence rates under regularization}
The forecast accuracy of both the ridge-regularized forecast \eqref{eq:ridgeforecast} and the random projection forecast \eqref{eq:rpforecast}, depends crucially on the choice of the regularization parameter $k$. The following theorem stipulates the conditions imposed on $k$ needed for consistency, as well as the resulting convergence rate.

\begin{theorem}\label{thm:1} Let \crefrange{assA1}{ass:varepsilon} hold.  If $k$ is such that as $(N,T)\rightarrow\infty$,  $\delta_{NT}^{-2}k\rightarrow 0$ and $\sqrt{\frac{N}{N^{\alpha}}}k^{-1}\rightarrow 0$, then
\begin{equation}\label{eq:thm1}
    \begin{split}
\hat{y}_{T+h|T}^{reg} - \bar{y}_{T+h|T} &=O_{p}\left(\frac{1}{N^{\alpha/2}}\right)+  O_{p}\left(\frac{1}{T^{1/2}}\right)+ O_{p}\left(\frac{N^{3/2}}{N^{3\alpha/2}T}\right)\\
& +O_{p}\left(\frac{N^{1/2}}{N^{3\alpha/2}}\right) +  O_{p}\left(\frac{N^2}{N^{2\alpha}T^{3/2}}\right)+O_{p}\left(\frac{N}{N^{2\alpha}T^{1/2}}\right)\\
& + O_{p}\left(\frac{N}{N^{\alpha}}k^{-1}\right)+ O_{p}\left(k \delta_{NT}^{-1}\left[\sqrt{\frac{N}{N^{\alpha}}\frac{1}{T}}+\sqrt{\frac{1}{N^{\alpha}}} + \frac{1}{N^{\alpha/4}T^{1/4}}\right]\right).
\end{split}
\end{equation}
\end{theorem}
The first two lines are identical to those obtained for the PCA-based forecast. The second line shows how the regularization parameter balances capturing the signal from the true factors while suppressing the noise coming from the idiosyncratic errors. Again, we specialize this result to three special cases considered before.
\begin{corollary}\label{corr3}
    Case 1. Strong loadings: $\alpha=1$. Set the regularization parameter as $k=O((N\delta_{NT}^2)^{1/4})$.
    \begin{equation}
        \hat{y}_{T+h|T}^{reg}- \bar{y}_{T+h|T}= O_{p}(\delta_{NT}^{-1}).
    \end{equation}
    Case 2. Weaker loadings: $\alpha<1$. $N\asymp T$. Set the regularization parameter as $k=O(N^{(3-\alpha)/4})$.
    \begin{equation}
        \hat{y}_{T+h|T}^{reg}- \bar{y}_{T+h|T}= O_{p}\left(\frac{1}{N^{\alpha/2}}\right) + O_{p}\left(\frac{1}{N^{(3\alpha-1)/4}}\right) .
    \end{equation}
    Case 3. Weaker loadings: $\alpha<1$. $T=O(N^{\gamma})$, $0<\gamma<1$. Set the regularization parameter as $k=O(N^{(1+2\gamma-\alpha)/4})$. Then,
    \begin{equation}
        \begin{split}
            \hat{y}_{T+h|T}^{reg} - \bar{y}_{T+h|T} &= O_{p}\left(\frac{1}{N^{\alpha/2}}\right) + O_{p}\left(\frac{1}{N^{\gamma/2}}\right) +O_{p}\left(\frac{1}{N^{(3\alpha-3+2\gamma)/4}}\right).
        \end{split}
    \end{equation}
\end{corollary}
\begin{figure}[t]
    \centering
    \caption{Convergence rates for PCA-based forecast and regularization-based forecasts}
    \label{fig:inequalities}
    \begin{subfigure}{\twosubf}
        \includegraphics[width=\textwidth]{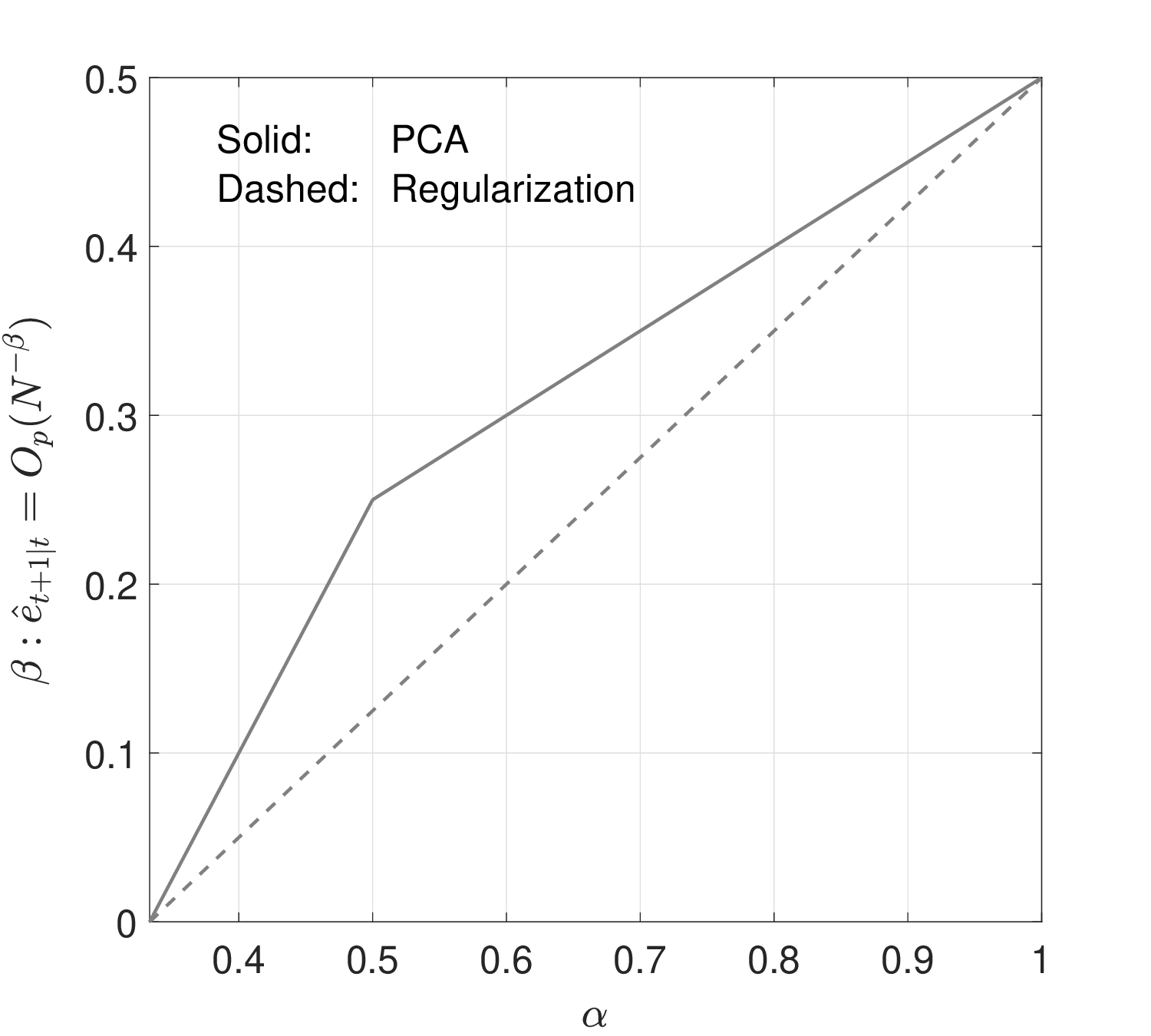}
    \end{subfigure} \quad
    \begin{subfigure}{\twosubf}
        \includegraphics[width=\textwidth]{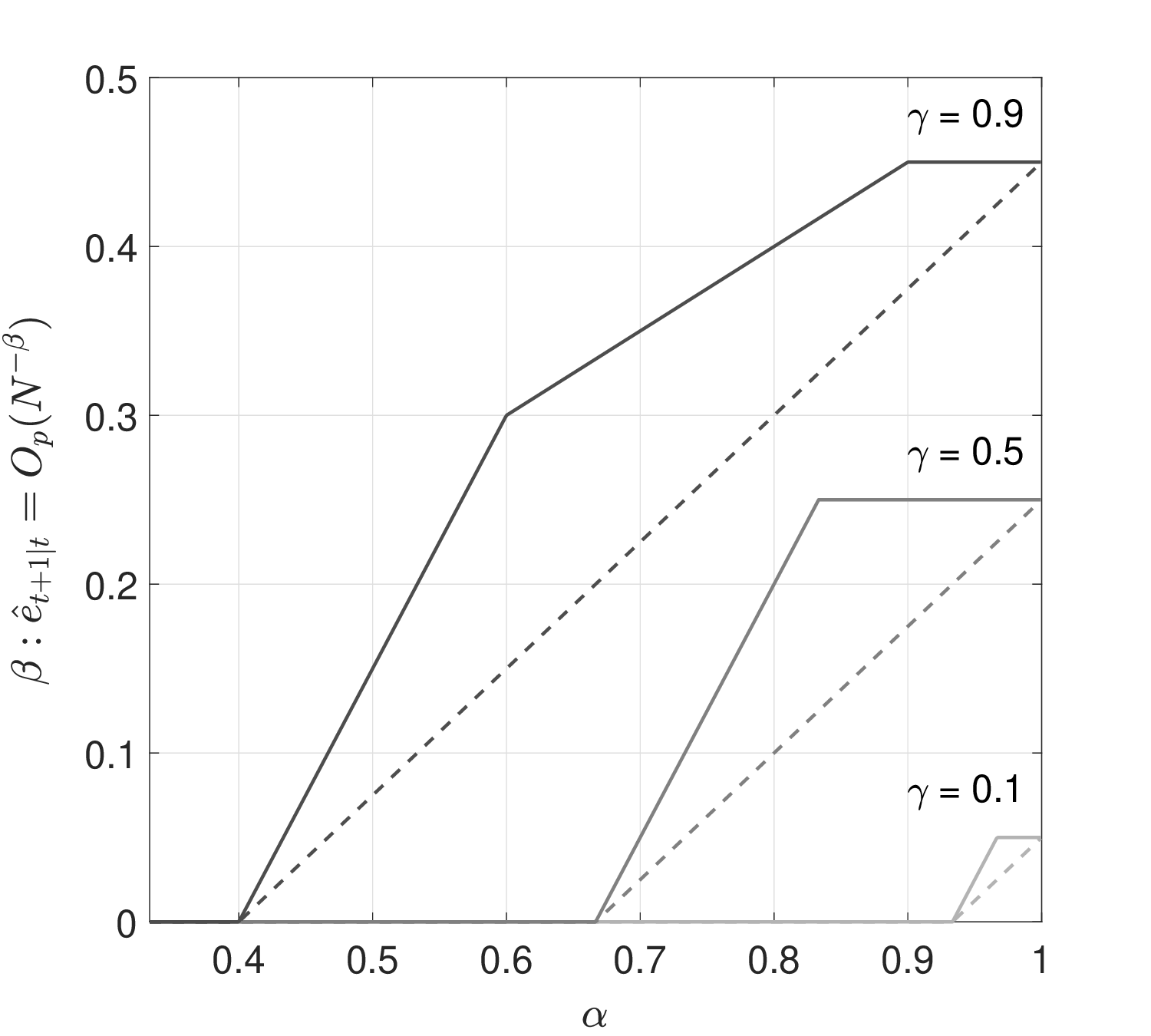}
    \end{subfigure}
    \floatexpl{The figure presents the convergence rate achieved for different values of the parameter $\alpha$ that governs the strength of the loadings. The solid line shows the convergence rate for PCA-based forecasts, the dashed line for ridge regression and random projection forecasts. The left panel shows the case where $N\asymp T$, the right panel the case where $T = O(N^{\gamma})$ for $\gamma\in \{0.1,  0.5, 0.9\}$.
    }
\end{figure}
In Case 1, we obtain the convergence rate $\delta_{NT}^{-2}$. This is identical to PCA and improves the convergence rate presented in \citet{de2008forecasting} for ridge regression. In Case 2
with weaker loadings and $N\asymp T$, \cref{corr3} shows that for consistency, it is again sufficient that $\alpha>1/3$. However, the second term under the optimal choice of the regularization parameter $k$ decays slower to zero compared to what is the case for PCA-based forecasts. \cref{fig:inequalities} plots the different convergence rates for the PCA-based forecast and the regularization-based forecasts as a function of the strength of the loadings. 

In Case 3, when $T = N^{\gamma}$ for $\gamma<1$, we reach a similar conclusion as for Case 2: the conditions under which the regularization-based forecasts are consistent are the same as for PCA, but regularization results in a lower convergence rate. The right panel of \cref{fig:inequalities} graphically shows the difference in the convergence rate for $\gamma=\{0.1,0.5,0.9\}$.

\subsection{Faster convergence rates under serially uncorrelated idiosyncratic errors}\label{subsec:improved}
The results of the previous section show that in general, we expect the regularization methods to suffer under weaker loadings. However, we now show that a substantial improvement over the convergence rates established in \cref{thm:0} and \cref{thm:1} can be obtained by introducing \cref{assA5},  under which the rows of the idiosyncratic error term are independent. In this case, we have the following.
\begin{theorem}\label{thm:2} Under \crefrange{assA1}{assA5}, the PCA-based forecast \eqref{eq:pcaforecast} satisfies
    \begin{equation}
        \begin{split}
            \hat{y}_{T+h|T}^{pca} -\bar{y}_{T+h|T}&=O_{p}\left(\frac{1}{N^{\alpha/2}}\right)+  O_{p}\left(\frac{1}{T^{1/2}}\right).
        \end{split}
    \end{equation}
\end{theorem}
We see that the condition that $\alpha>1/3$ is no longer necessary. What is still required is \cref{assA4}. Case 3, where $T = O(N^{\gamma})$, therefore requires $\alpha+\gamma>1$.

The results for ridge regression and random projection forecasts can be similarly improved to the following.

\begin{figure}[t]
    \centering
    \caption{Convergence rates under time independent noise and $T=O(N^{\gamma})$}
    \label{fig:inequalitiesimproved}
    \includegraphics[width=0.8\textwidth]{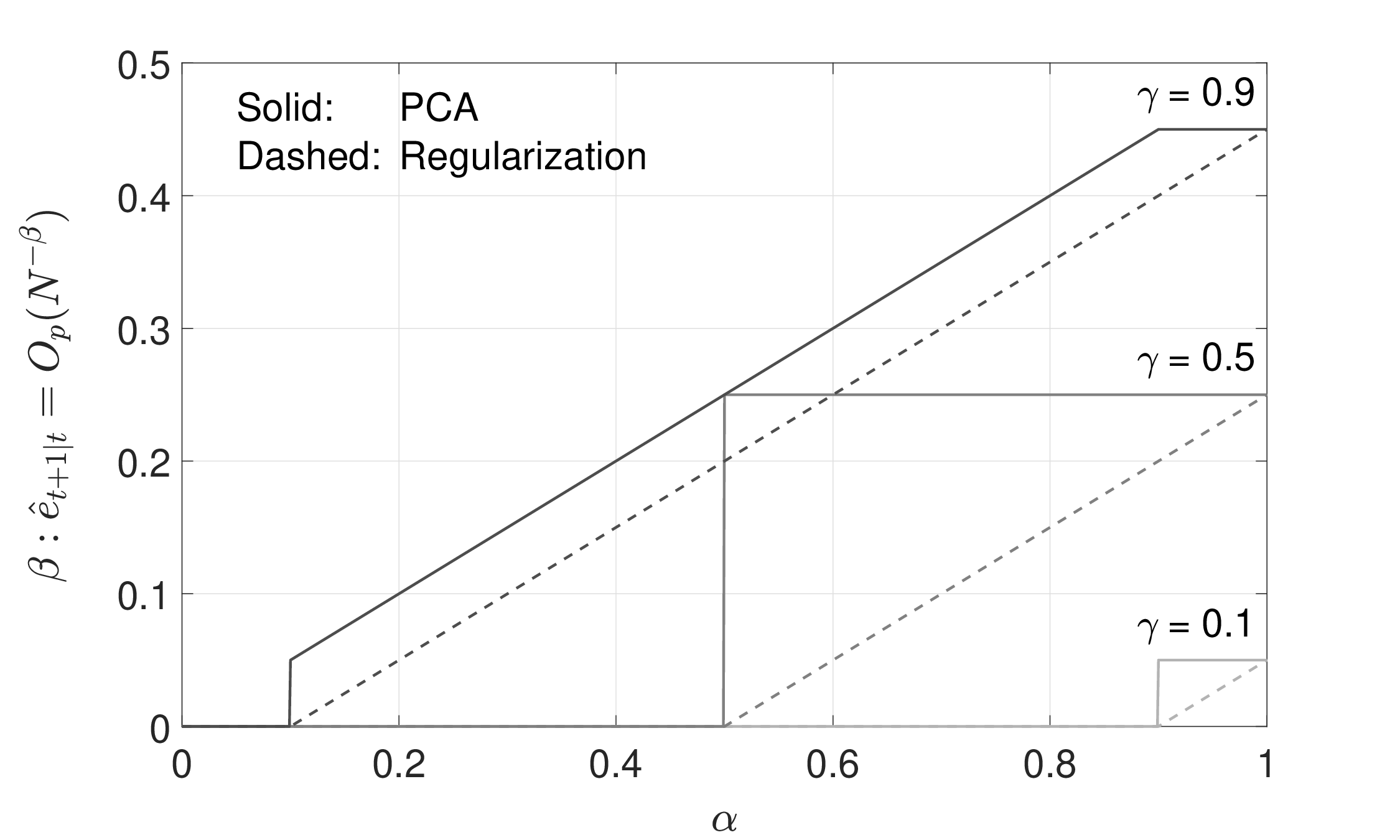}
    \floatexpl{The figure presents the convergence rate achieved for different values of the parameter $\alpha$ that governs the strength of the loadings and the parameter $\gamma$ that determines the expansion rate of $T/N^{\gamma}= O(1)$. 
    }
\end{figure}

\begin{theorem}\label{thm:3} Under \crefrange{assA1}{assA5}, and setting  $k$ such that as $(N,T)\rightarrow\infty$,  $\delta_{NT}^{-2}k\rightarrow 0$ and $\sqrt{\frac{N}{N^{\alpha}}}k^{-1}\rightarrow 0$, the regularization-based forecast \eqref{eq:rpforecast} satisfies
    \begin{equation}
            \hat{y}_{T+h|T}^{reg} -  \bar{y}_{T+h|T}=O_{p}\left(\frac{1}{N^{\alpha/2}}\right)+  O_{p}\left(\frac{1}{T^{1/2}}\right)+ O_{p}\left(\frac{N}{N^{\alpha}}k^{-1}\right)+ O_{p}\left(k\cdot \delta_{NT}^{-2}\right).
    \end{equation}
\end{theorem}
For Case 1 with $k = O(\delta_{NT})$ and Case 2 with $k=O(N^{1-\alpha/2})$, we now obtain the same result as presented for PCA in \Cref{thm:2},
\begin{equation}
    \begin{split}
        \hat{y}_{T+h|T} - \bar{y}_{T+h|T} &=O_{p}\left(\frac{1}{N^{\alpha/2}}\right)+  O_{p}\left(\frac{1}{T^{1/2}}\right).
    \end{split}
\end{equation}
However, in Case 3, when $T = N^{\gamma}$ for $0<\gamma<1$, and $k=O(N^{(1-\alpha+\gamma)/2})$, we obtain
\begin{equation}
    \begin{split}
        \hat{y}_{T+h|T}^{reg} - \bar{y}_{T+h|T} &=  O_{p}\left(\frac{1}{N^{\alpha/2}}\right) + O_{p}\left(\frac{1}{N^{\gamma/2}}\right) +O_{p}\left(\frac{1}{N^{(\alpha+\gamma-1)/2}}\right).
    \end{split}
\end{equation}
The requirement for consistency is $\alpha+\gamma>1$ as it is for PCA. However, in some parts of the parameter space, especially if both $\alpha$ and $\gamma$ are substantially below 1, the convergence rate is lower compared to PCA. The convergence rates are shown in \cref{fig:inequalitiesimproved} for different values of $\gamma$ and $\alpha$. 

\section{Simulations}\label{sec:simulations}
We consider the model \eqref{eq:factorequation} and \eqref{eq:forecastequation} with $r=2$ factors. The elements of the factors $\bs F$ are independent standard normal random variables. To tune the strength of the loadings, we generate $[\bs\Lambda]_{ij} = \sqrt{\frac{N^{\alpha}}{N}}Z_{ij}$ with $Z_{ij}$ standard normal independent over $i=1,\ldots,N$ and $j=1,\ldots,r$. The parameter $\alpha$ will be varied over $\{0.5,0.75,1\}$ where $\alpha=1$ corresponds to a setting with strong factors. We first consider a setting with the idiosyncratic noise generated as $\bs e_{ti}\sim N(0,r)$, independently across $t=1,\ldots,T$ and $i=1,\ldots,N$. We subsequently consider a setting with serial correlation in the idiosyncratic noise. 

In the forecast equation \eqref{eq:forecastequation}, the parameter $\bs \gamma=(1,1)'$ and $\varepsilon_{t}\sim N(0,r)$ independently over $t$. We consider one-step ahead forecasts and set $\varepsilon_{T+1}=0$ as this is a noise term that we cannot forecast. We generate PCA-based forecasts assuming that the number of factors is known. For ridge regression and random projections, we pick $k$ ex post to minimize the mean squared forecast error. We compare the forecasts in terms of their Mean Squared Forecast Error (MSFE) defined as $\mathbb{E}[(\hat{y}_{T+h|T}-\bar{y}_{T+h|T})^2]$. Finally, we vary $N$ and $T$ over the grid $\{100, 200, 300, 400, 500\}$. Results are averaged over 5,000 realizations of the data generating process.

\begin{figure}[t!]
    \centering
    \caption{Mean squared error and regularization parameter ($N=T$)}
    \label{fig:MC1}
   \includegraphics[width=\textwidth]{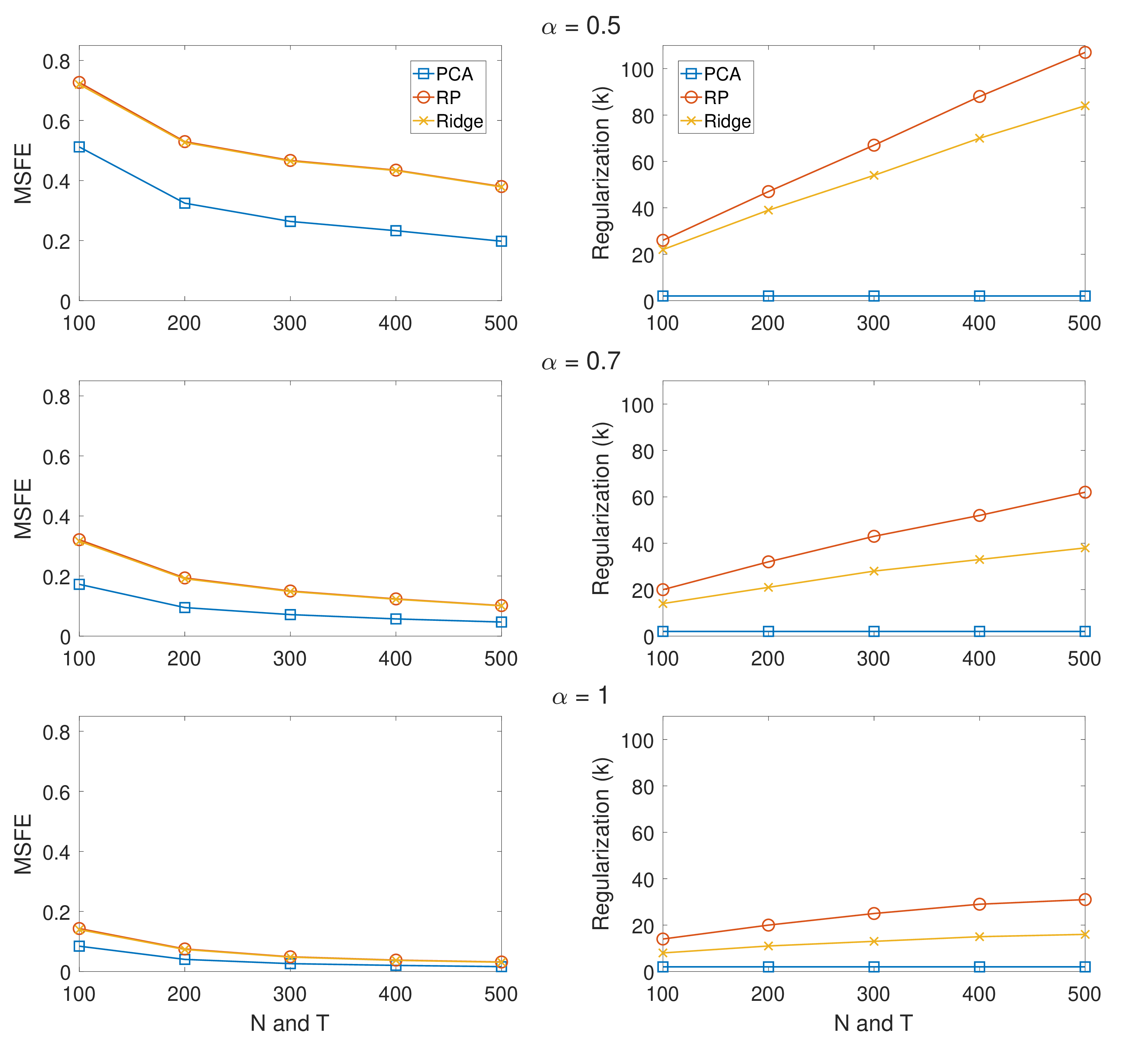}
   \floatexpl{The figure presents the Mean Squared Forecast Error (MSFE)  $\hat{\mathbb{E}}[(\hat{y}_{T+1|T}-\bar{y}_{T+1|T})^2]$ in the left panel, with $\hat{\mathbb{E}}$ indicating the average over 5,000 Monte Carlo simulations. The data generating process accords with \eqref{eq:factorequation} and \eqref{eq:forecastequation} with $r=2$ factors, and the elements of the factors, loadings and idiosyncratic noise terms are all independent standard normals. We compare the PCA-based forecast, the ridge regression forecast and the random projection forecast. For the latter two, the choice of $k$ is shown in the right panel. We set $T=N$ and consider $\alpha=0.5$ (upper panel), $\alpha=0.75$ (middle panel) and $\alpha=1$ (lower panel). 
   }
\end{figure}

\subsection{Independent idiosyncratic noise}
The left panel of \cref{fig:MC1} shows the one-step ahead MSFE for PCA-based forecasts as well as ridge and random projection forecasts when we set $T=N$. The right panel of the figure displays the optimal choice of the regularization parameter for the latter two methods. We observe the following: regardless of the value of $\alpha$ and the values of $N$ and $T$, PCA-based forecasts offer the lowest MSFE. There is no observable difference between ridge and random projection forecasts in terms of the MSFE. We also see that as the loadings become weaker, the difference between PCA and the regularization methods appears to increase. With regard to the regularization parameter, we see that it increases with increasing $(N,T)$ and with weaker loadings.

\begin{figure}[t!]
   \centering
   \caption{MSFE and regularization parameter (varying $T$ and $N$)\vspace{0.2cm}}
   \label{fig:MC2}
   \begin{subfigure}{\textwidth}
   \caption{Weak loadings $(\alpha=0.7)$.\vspace{-0.2cm}}
   \includegraphics[width=\textwidth]{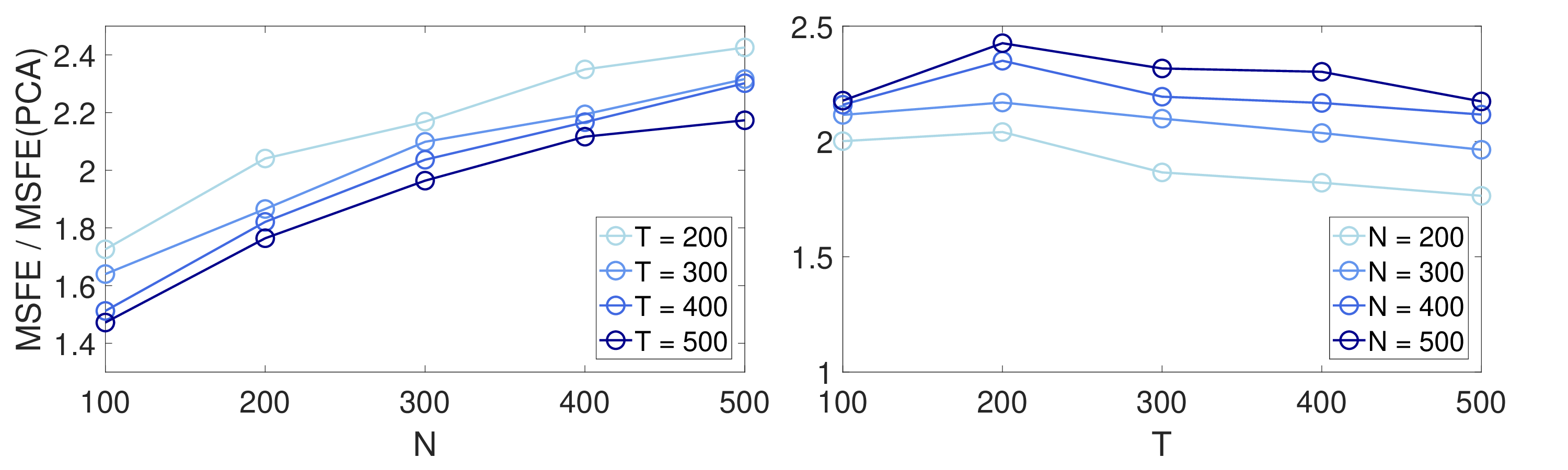}
   \end{subfigure}
\begin{subfigure}{\textwidth}   
\caption{Strong loadings $(\alpha=1)$.\vspace{-0.2cm}}
    \includegraphics[width=\textwidth]{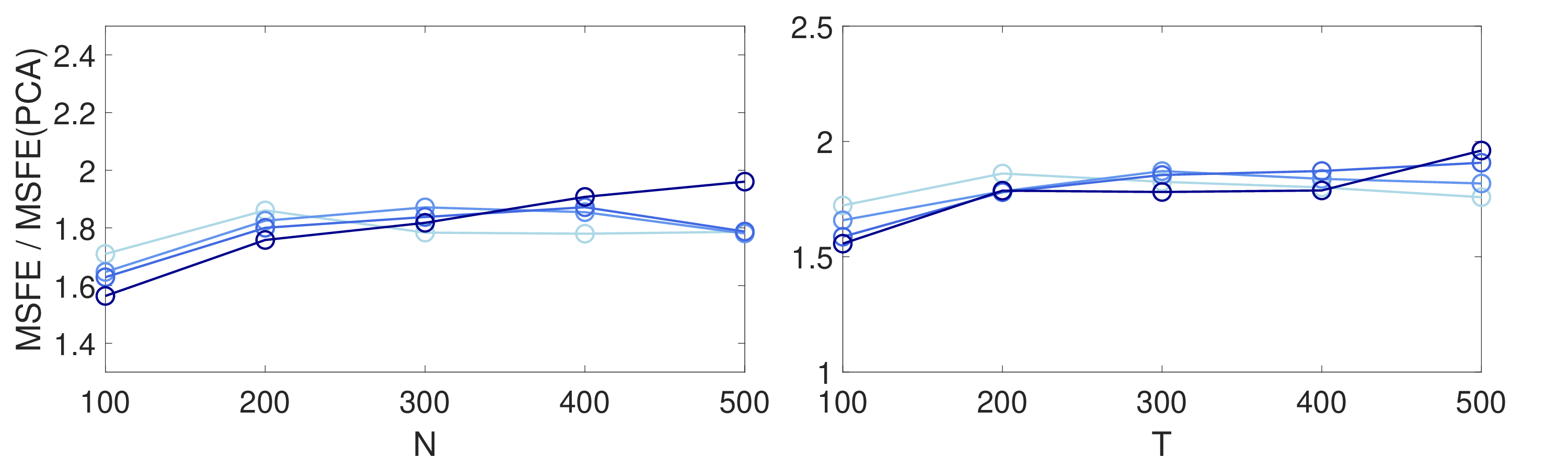}
    \end{subfigure}
   \floatexpl{The figure presents the Mean Squared Forecast Error (MSFE) $\hat{\mathbb{E}}[(\hat{y}_{T+1|T}-\bar{y}_{T+1|T})^2]$ for ridge forecasts relative to the MSFE of the PCA-based forecasts. Top panel has weak loadings with $\alpha = 0.75$, bottom panel has strong loadings with $\alpha = 1$. In the left panel, we fix $T$ at $\{200,300,400,500\}$ and vary $N$ on the $x$-axis. In the right panel, we fix $N$ at $\{200,300,400,500\}$ and vary $T$ on the $x$-axis. For further details, see the note following \cref{fig:MC1}.
   }
\end{figure}

\cref{fig:MC2} shows the MSFE of ridge regression relative to PCA-based forecasts. We omit the random projection forecasts, because the accuracy is identical to the ridge regression forecasts. 
In the left panel of \cref{fig:MC2}, we fix $T$ and vary $N$ on the $x$-axis. In the right panel, we fix $N$ and vary $T$ on the $x$-axis. The upper panel has weak loadings with $\alpha=0.75$, while the lower panel has strong loadings with $\alpha=1$. 
The most important finding is that for strong loadings, the relative performance between regularization-based methods and PCA stabilizes as either $N$ or $T$ increases. This confirms that the convergence rates for these methods are identical under strong loadings. Under weaker loadings (upper panel), we see that the relative MSFE stabilizes with increasing $T$ (right panel), yet it continues to increase with larger $N$ (left panel). This confirms the theoretical results from \cref{subsec:improved} that state that under weaker loadings, the regularization methods may attain a lower convergence rate when $N$ is increasing at a faster rate compared to $T$.

\subsection{Serially correlated idiosyncratic errors}
We change the setup by introducing serial correlation in the idiosyncratic component of the predictors. To be precise, we consider $E_{ti} = \rho E_{t-1,i} + v_{t-1,i}$ with $v_{ti}\sim N(0,1-\rho^2)$ independent over $t$ and $i$, $E_{1i}=v_{1i}\sim N(0,1)$ and $\rho=0.7$.

\begin{figure}[t!]
   \centering
   \caption{MSFE and regularization parameter with serially correlated errors}
   \label{fig:MC3}
   \includegraphics[width=\textwidth]{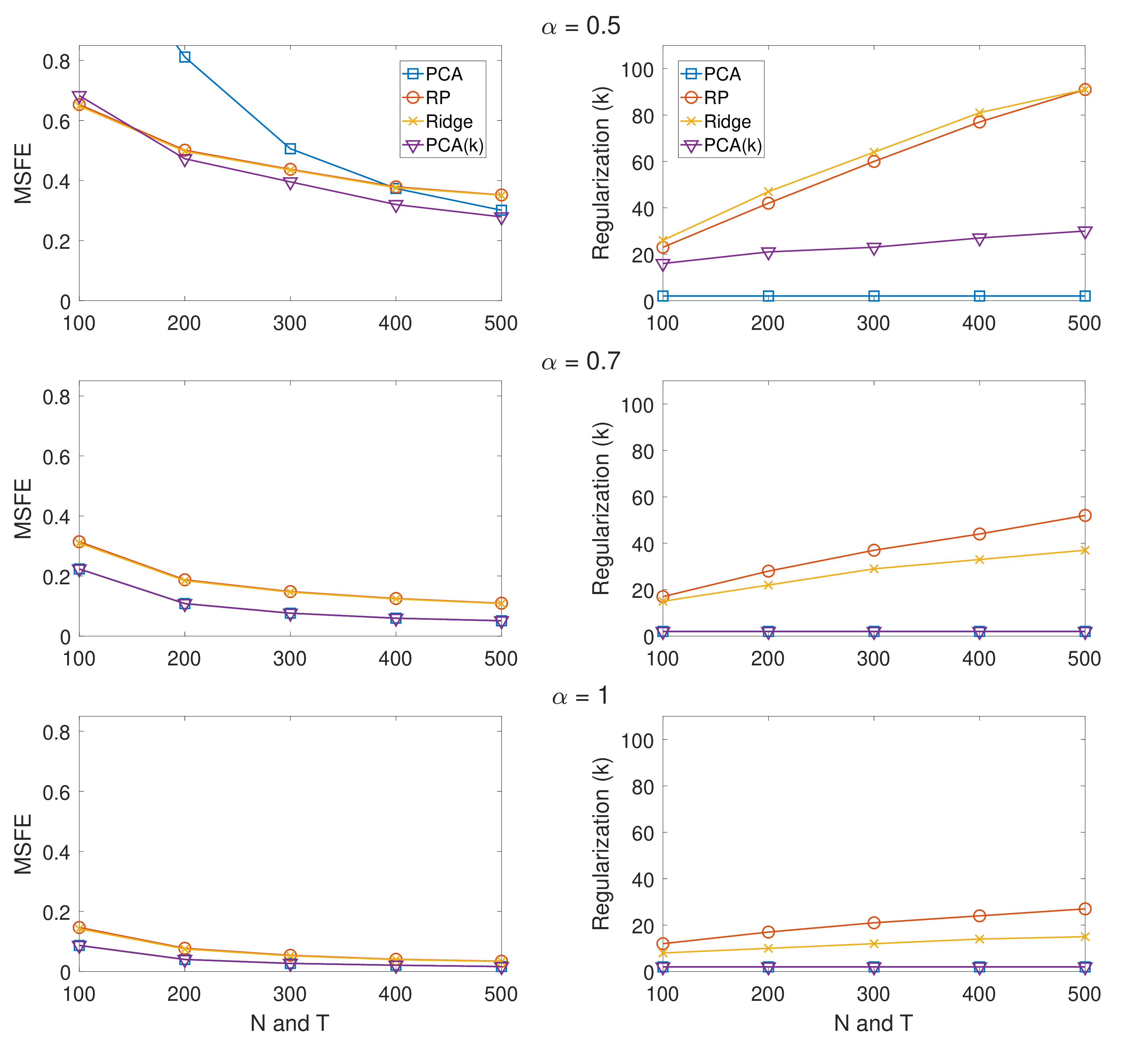}
   \floatexpl{Details are in the note following \cref{fig:MC1}, with the addition that the idiosyncratic noise follows an AR(1) process with autocorrelation parameter $\rho=0.7$. PCA(k) denotes a PCA-based forecast with ex post optimal selection of the number of factors.
   }
\end{figure}

\cref{fig:MC3} shows again the MSFE and the optimal choice for the regularization parameter, akin to \cref{fig:MC1}. There are two important observations. First, we see a noticeable deterioration in the performance of PCA-based forecasts under weak factors and small sample sizes. There is hardly any change in the MSFE for the ridge or random projection forecasts, and we only find a slightly smaller optimal value of $k$.
The deterioration of PCA can be explained intuitively by the fact that the eigenvalues are no longer clearly separated from the largest eigenvalue of the idiosyncratic errors. Suppose that the three largest eigenvalues are of the same order of magnitude. Because the autocorrelation increases the maximum eigenvalue of the covariance matrix of the idiosyncratic noise, it may be the case that the second largest eigenvalue is associated with a singular vector that is not related to the unobserved factors, while the third largest eigenvalue is associated with the unobserved factors. In this case, PCA would miss the relevant signal as it hard thresholds the eigenvalues at the supposed true number of factors $r$. The regularization-based methods, on the other hand, will apply comparable soft-thresholding to the largest three eigenvalues, maintaining the majority of the signal and thereby improving over PCA. 

\cref{fig:MC3} also adds \textit{PCA(k)}, which selects the optimal number of factors ex post. We see that indeed for weak loadings, the optimal number of factors is much larger than the nominal number of factors. Moreover, choosing the optimal number of factors largely mitigates the deterioration of the PCA-based forecasts. For small sample sizes, and weak factors, the regularization-based methods continue to offer more accurate forecasts. The second important observation is however that when $\alpha=0.5$ and we use \textit{PCA(k)}, we see that the MSFE for PCA converges faster to zero relative to random projections and ridge. This confirms the theoretical results in \cref{corr3} Case 2 versus \cref{corr2} Case 2.

\section{Application: forecast accuracy on monthly and quarterly macroeconomic data sets}
To confirm the theoretical results in practice, we setup a forecasting exercise to assess the impact of the ratio $T/N$ on forecasts from PCA, ridge and random projections in an empirical setting. The theory suggests that as $T/N$ decreases, PCA becomes relatively more accurate. However, the simulations also show that in small samples, the regularization methods may yield more accurate forecasts. Our goal is therefore analyse prediction accuracy on variables from the FRED-MD \citep{mccracken2015fred} and FRED-QD \citep{mccracken2020fred}. These data sets are large cross-sections of macroeconomic and financial variables and have been studied using PCA and random projections \citep{boot2019forecasting}. We select vintage 2024-01 for both data sets.

For each variable, forecasts are generated from the model
\begin{equation}
    y_{t+h} = \bs w_t' \bs \beta_w + \bs x_t' \bs R \bs\beta_x + \varepsilon_{t+h},
\end{equation}
where the $N_w \times 1$ vector $\bs w_t$ are the first four lags of $y_t$ and an intercept, which are always included. The $N_x \times 1$ vector $\bs x_t$ are the fifth and sixth lag of the target variables, and all the other variables. The $N_x \times k$ matrix $\bs R$ with $k < N_x$ reduces the dimension of the regressors. 

The sample runs from 1959M1 (or 1959Q1) up to the end of 2023. Forecasting starts in the first period of 1970, i.e. 1970M1 for the monthly set and 1970Q1 for quarterly data. Out-of-sample evaluation begins in 1980M1 or 1980Q1, after a 10-year burn-in to select the hyperparameters that minimize the squared forecast error. So, the number of factors, the subspace dimension, and the ridge penalty are optimized over an expanding window that starts in 1970. The maximum number of factors in PCA and the subspace dimension in RP are set to 50. Random projection forecasts are averaged over 1,000 draws. For ridge, we set up a grid for the log of the penalty parameter that runs from -14.7 to 15, with increments of 0.3, following \cite{boot2019forecasting}.

We conduct two experiments to assess the impact of the ratio $T/N_x$, changing the length or frequency of the estimation sample while keeping the number of regressors, $N_x$, fixed.

\begin{figure}[t]
	\caption{Percentage of times MSFE is smaller for PCA \label{fig:app:PCAwins}}
	\centering
	\includegraphics[width=0.7\textwidth,trim=0 0.5cm 0 0,clip]{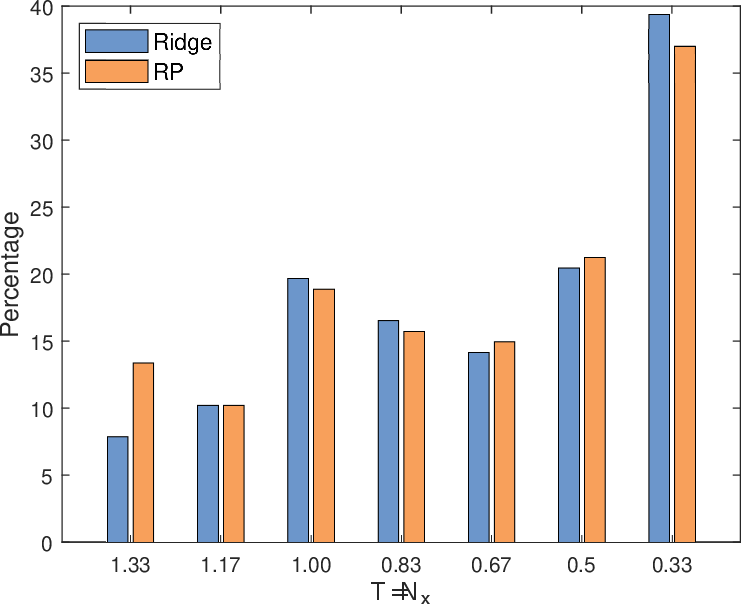}\newline
        $T/N_x\hspace{0.75cm}$
    \floatexpl{The figure presents the percentage of times that forecasts from PCA has a smaller mean squared forecast error (MSFE) loss compared to forecasts using ridge (blue) or random projections (RP, orange) out of 127 variables from the FRED-MD. The fraction $T/N_x$ changes from varying moving window lengths while $N_x$ is fixed. The evaluation sample is 1980M1--2023M12.
	}
\end{figure}

\subsection{Moving window size}
First, we vary the size of the moving window, $T$, from $\tfrac{1}{3}N_x$ to $\tfrac{4}{3}N_x$. The advantage of comparing across moving windows is that it controls the ratio $T/N_x$ as it is fixed over the evaluation sample. Monthly data allows for a wider range of moving windows and a sufficient number of forecasts for evaluation.

\Cref{fig:app:PCAwins} presents the percentage of times that PCA-based forecasts achieve a smaller MSFE compared to ridge or random projection forecasts.
The percentages are below 50 percent for all window sizes, indicating that ridge and random projections yield on average more accurate forecasts than PCA in this dataset. For shorter moving window lengths, the forecasts obtained by PCA improve in terms of MSFE relative to ridge and random projections. With a window length of $T= \tfrac{7}{6} N_x$, PCA yields better forecasts in about 10 percent of the cases, while PCA-based forecasts are more accurate than ridge and random projections for close to 40 percent of the variables when $T=\tfrac{1}{3}N_x$. The increase in relative performance between longer and shorter windows is not monotonic, but there is a clear pattern indicating that ridge and random projections yield relatively less accurate forecasts than PCA at shorter windows.

\begin{figure}[t]
    \caption{DM test statistics \label{fig:app:DMstats}}
    \centering
    \begin{subfigure}{\twosubf}
	\caption{Ridge}
	\includegraphics[width=\textwidth]{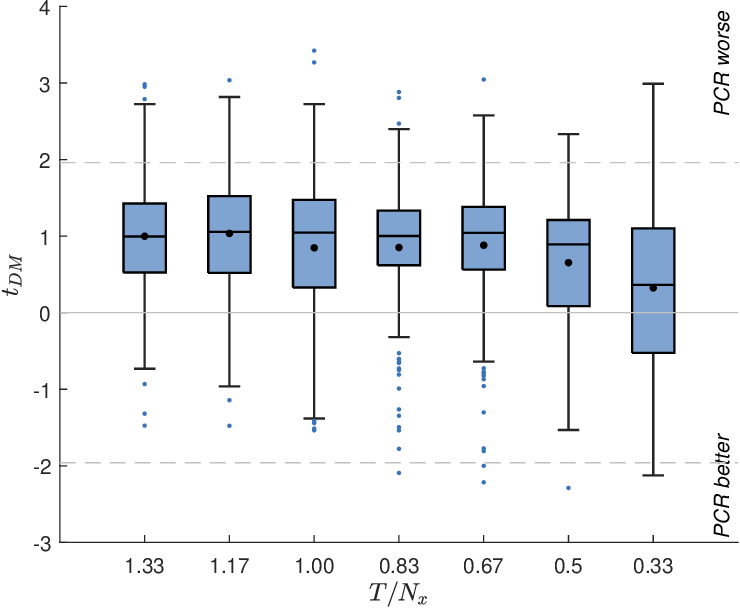}
    \end{subfigure} \quad
    \begin{subfigure}{\twosubf}
	\caption{Random projections}
	\includegraphics[width=\textwidth]{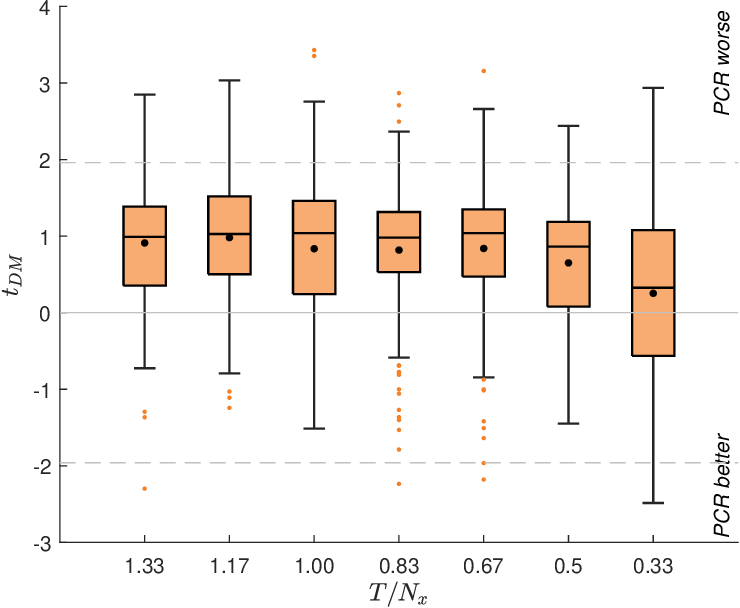}
    \end{subfigure}
    \floatexpl{The figure presents box plots of DM test statistics for the difference in squared error loss between ridge (left panel) or random projections (right panel) forecasts compared to PCA-based forecasts. The fraction $T/N_x$ changes from varying moving window lengths while $N_x$ is fixed. Black dots represent the mean DM statistic. Light gray dashed lines correspond to the two-sided 5\% critical value, assuming a standard normal distribution. The evaluation sample is 1980M1--2023M12.}
\end{figure}

\Cref{fig:app:DMstats} presents the distribution of Diebold-Mariano (DM) test statistics \citep{diebold1995comparing} computed on the difference between the squared error loss from regularization- and PCA-based forecasts. Positive values correspond to more accurate forecasts by ridge or random projections. The distribution of DM statistics is shifted downwards for shorter windows relative to the distribution for longer windows. DM statistics are more often negative with shorter windows, see also \Cref{fig:app:PCAwins}. When forecasting CPI: Durables, PCA even produces significantly more accurate forecasts than ridge or random projections with windows shorter than $T=\tfrac{5}{6}N_x$, while the DM statistics are positive with windows larger than $T=N_x$.

\begin{figure}[t]
	\caption{Change in DM test statistic \label{fig:app:DeltaDMstats}}
	\centering
	\begin{subfigure}{\textwidth}
        \caption{Ridge}
        \includegraphics[width=\textwidth]{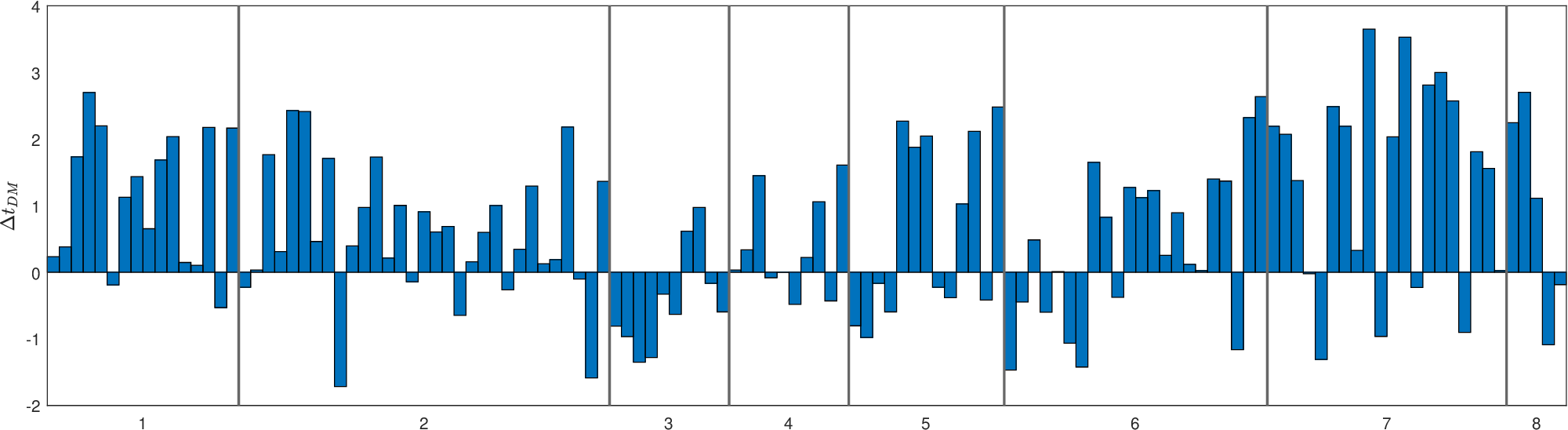}
    \end{subfigure} \\
    \begin{subfigure}{\textwidth}
        \caption{Random projections}
        \includegraphics[width=\textwidth]{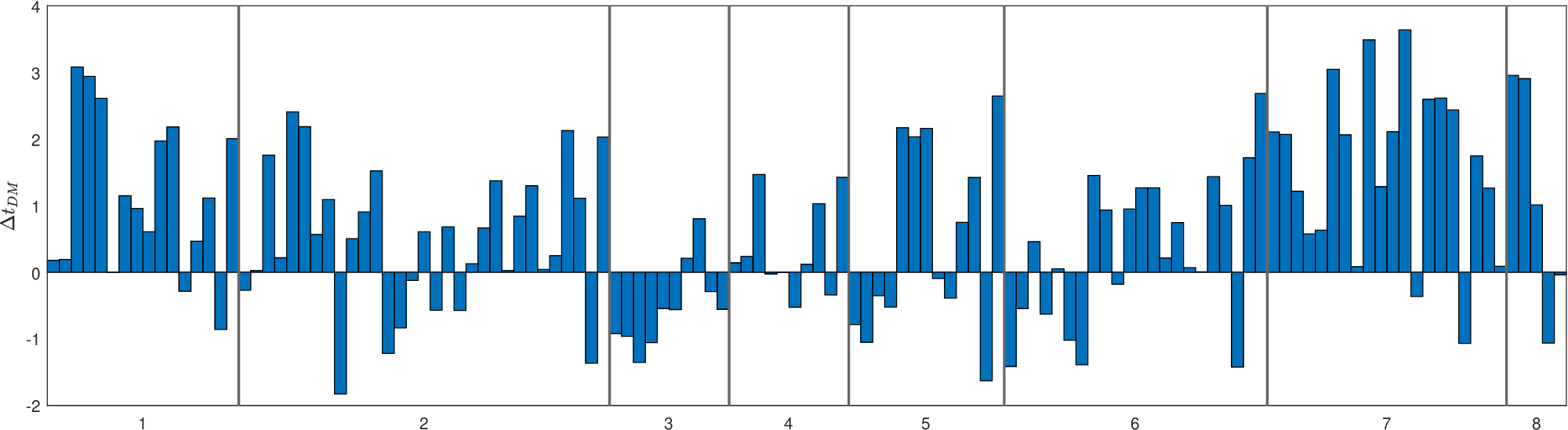}
    \end{subfigure}
	\floatexpl{The figure presents change in DM test statistic across estimation windows for each variable in the FRED-MD. The $y$-axis shows the DM test statistics from the longest estimation window ($T=4/3N_x$) minus the DM test statistic from the shortest window ($T=1/3N_x$), where the DM test statistic is computed on the squared forecast loss differential between ridge (top panel) or random projections (bottom panel) versus PCA.
        The evaluation sample is 1980M1--2023M12. Numbers on the $x$-axis denote the FRED-MD groups: (1) Output and income; (2) Labor market; (3) Housing; (4) Consumption, orders, and inventories; (5) Money and credit; (6) Interest and exchange rates; (7) Prices; (8) Stock market.
	}
\end{figure}

\Cref{fig:app:DeltaDMstats} presents the difference in DM test statistics between the largest window ($T=\tfrac{4}{3}N_x$) and the shortest window ($T=\tfrac{1}{3}N_x$), for each of the variables in the FRED-MD. Positive values indicate a shift of relative improvement for PCA compared to ridge or random projections with the shortest window compared to the larger window. The shift in the distribution of DM test statistics is not driven by a particular set of variables because the change is positive for most variables. The shift is particularly large for variables in categories 1 (Output and income) and 7 (Prices) and results in a change in the preferred model. For example, the DM statistic from comparing ridge (random projections) with PCA in the case of IP: Final Products and Nonindustrial Supplies is 1.30 (1.31) with $T=\tfrac{4}{3}N_x$, while it is -1.40 (-1.63) with $T=\tfrac{4}{3}N_x$.
As an exception, PCA perform relatively worse for variables in category 3 (Housing).

\subsection{Frequency}
As a second experiment, we compare the FRED-MD and FRED-QD data sets using an estimation window covering the same calendar period but at different frequencies, thus varying $T/N_x$. We employ a moving window of 10 years, such that the number of observations is $T=120$ for monthly and $T=40$ for quarterly data. We consider the subset of 102 variables that are in both the FRED-MD and the FRED-QD, see Appendix \ref{app:sec:innersubset}.

Monthly observations are aggregated to the quarterly level and monthly forecasts are at a $h=3$ month horizon. The quarterly forecasts are at a $h=1$ quarter horizon. Hence, the observations and forecasts measure the same entity across the data sets, only at different frequencies.

\begin{figure}[t]
    \caption{DM test statistics across data frequency \label{fig:app:DMstats-freq}}
    \centering
    \begin{subfigure}{\twosubf}
        \caption{Ridge}
        \includegraphics[width=\textwidth]{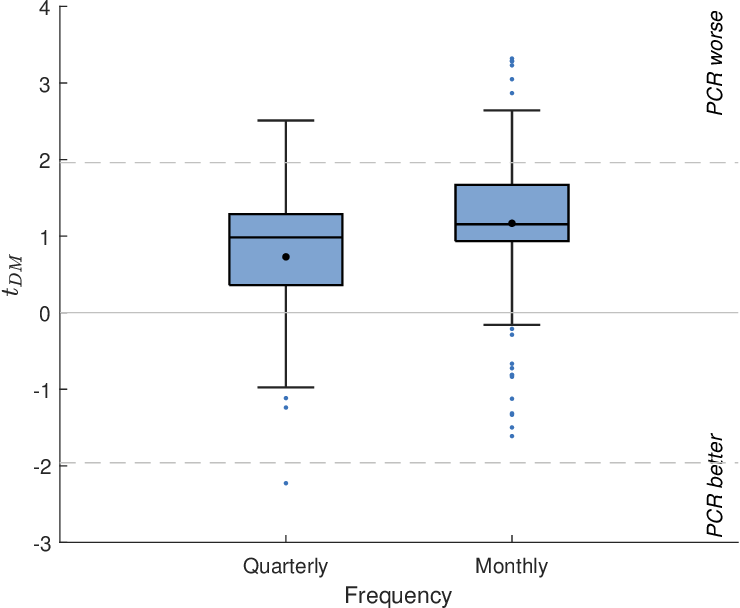}
    \end{subfigure} \quad
    \begin{subfigure}{\twosubf}
        \caption{Random projections}
        \includegraphics[width=\textwidth]{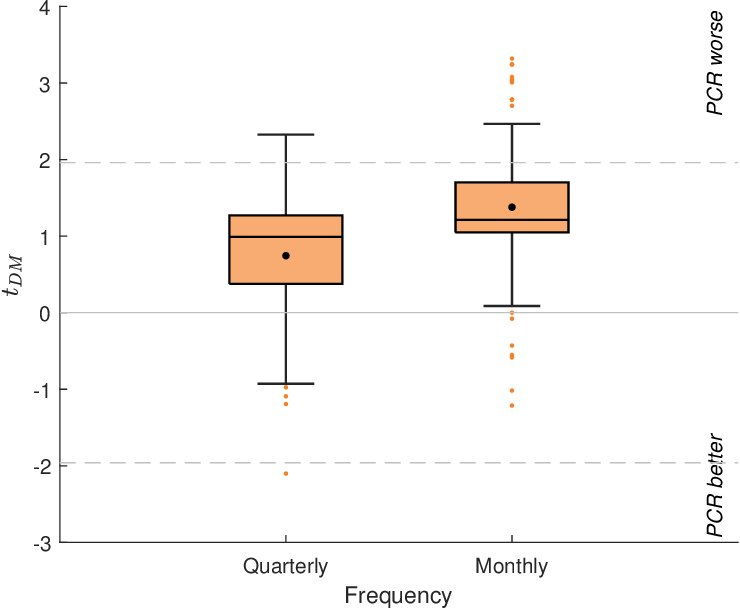}
    \end{subfigure}
    \floatexpl{The figure presents box plots of DM test statistics for the difference in squared error loss between ridge (left panel) or random projections (right panel) forecasts compared to PCA-based forecasts. The fraction $T/N_x$ changes from varying the frequency while $N_x$ is fixed. Black dots represent the mean DM statistic. Light gray dashed lines correspond to the two-sided 5\% critical value, assuming a standard normal distribution. The evaluation sample is 1980M1--2023M12.
    }
\end{figure}

In 19.6 percent of the cases, PCA has a smaller MSFE than ridge in the quarterly data set and only in 12.7 percent of the cases when using monthly data. Similarly, PCA leads to more accurate forecasts than random projections with 18.6 percent of the variables for quarterly and 5.9 percent for monthly data.

This corresponds to a shift in DM statistics between using data at the quarterly or monthly frequency, for both ridge and random projections, see \Cref{fig:app:DMstats-freq}. Using quarterly data results in DM statistics that are smaller than when using monthly data, implying that forecasts are preferred more often when using quarterly data. The change of the DM statistic can be substantial and lead to a different conclusion on which model yields more accurate forecasts. For example, the DM statistic for ridge (random projections) versus PCA is 3.32 (3.24) based on monthly data and -2.23  (-2.10) on quarterly data when forecasting the Switzerland / U.S. Foreign Exchange Rate.

Hence, the empirical findings are consistent between both experiments.
Ridge and random projections generally yield better forecasts in the FRED-MD and FRED-QD than PCA. However, when the estimation sample becomes smaller compared to the cross-sectional size, i.e. the fraction $T/N_x$ decreases, the accuracy of PCA-based forecasts improves relative to the accuracy of regularization-based forecasts.

\section{Conclusion}
We analyze the convergence rates of PCA-based forecasts in the diffusion index forecast model with weaker loadings. We compare these rates with forecasts based on ridge regression and random projections. We find that under strong loadings, all forecasts attain the same convergence rate. Under weaker loadings all forecasts are consistent under the same assumptions on the loading strength, but ridge regression and random projection forecasts converge more slowly relative to PCA when the cross-section dimension is large relative to the number of time periods. We confirm these results in simulations and an empirical study based on the FRED-MD and FRED-QD data sets. These numerical results also show that in small samples, regularization-based methods may be more robust to serial correlation in the idiosyncratic errors. 

\ifthenelse{\boolean{BLIND}}{}
{
\paragraph{Funding details}
Tom Boot acknowledges financial support by the Dutch Research Council (NWO) as part of grant VI.Veni.201E.11.
}

\paragraph{Disclosure statement} There are no competing interests to declare. 

\bibliographystyle{apalike}
\bibliography{literature}

\begin{thebibliography}{}

\bibitem[Achlioptas, 2003]{achlioptas2003database}
Achlioptas, D. (2003).
\newblock Database-friendly random projections: {J}ohnson-{L}indenstrauss with
  binary coins.
\newblock {\em Journal of Computer and System Sciences}, 66(4):671--687.

\bibitem[Ahn and Horenstein, 2013]{ahn2013eigenvalue}
Ahn, S.~C. and Horenstein, A.~R. (2013).
\newblock Eigenvalue ratio test for the number of factors.
\newblock {\em Econometrica}, 81(3):1203--1227.

\bibitem[Bai and Ng, 2006]{bai2006confidence}
Bai, J. and Ng, S. (2006).
\newblock Confidence intervals for diffusion index forecasts and inference for
  factor-augmented regressions.
\newblock {\em Econometrica}, 74(4):1133--1150.

\bibitem[Bai and Ng, 2023]{bai2023approximate}
Bai, J. and Ng, S. (2023).
\newblock Approximate factor models with weaker loadings.
\newblock {\em Journal of Econometrics}, 235(2):1893--1916.

\bibitem[Boot and Nibbering, 2019]{boot2019forecasting}
Boot, T. and Nibbering, D. (2019).
\newblock Forecasting using random subspace methods.
\newblock {\em Journal of Econometrics}, 209(2):391--406.

\bibitem[Brinkhuis et~al., 2005]{brinkhuis2005matrix}
Brinkhuis, J., Luo, Z.-Q., and Zhang, S. (2005).
\newblock Matrix convex functions with applications to weighted centers for
  semidefinite programming.
\newblock Technical Report No. EI 2005-38, Econometric Institute, Erasmus
  University Rotterdam.

\bibitem[Carrasco and Rossi, 2016]{carrasco2016sample}
Carrasco, M. and Rossi, B. (2016).
\newblock In-sample inference and forecasting in misspecified factor models.
\newblock {\em Journal of Business \& Economic Statistics}, 34(3):313--338.

\bibitem[Chiong and Shum, 2019]{chiong2016random}
Chiong, K.~X. and Shum, M. (2019).
\newblock Random projection estimation of discrete-choice models with large
  choice sets.
\newblock {\em Management Science}, 65(1):256--271.

\bibitem[De~Mol et~al., 2008]{de2008forecasting}
De~Mol, C., Giannone, D., and Reichlin, L. (2008).
\newblock Forecasting using a large number of predictors: Is {B}ayesian
  shrinkage a valid alternative to principal components?
\newblock {\em Journal of Econometrics}, 146(2):318--328.

\bibitem[De~Mol et~al., 2024]{de2024asymptotic}
De~Mol, C., Giannone, D., and Reichlin, L. (2024).
\newblock The asymptotic equivalence of ridge and principal component
  regression with many predictors.
\newblock {\em Econometrics and Statistics}.

\bibitem[Diebold and Mariano, 1995]{diebold1995comparing}
Diebold, F.~X. and Mariano, R.~S. (1995).
\newblock Comparing predictive accuracy.
\newblock {\em Journal of Business \& Economic Statistics}, 13(3):253--263.

\bibitem[Fan and Liao, 2022]{fan2020learning}
Fan, J. and Liao, Y. (2022).
\newblock Learning latent factors from diversified projections and its
  applications to over-estimated and weak factors.
\newblock {\em Journal of the American Statistical Association},
  117(538):909--924.

\bibitem[He, 2023]{he2024ridge}
He, Y. (2023).
\newblock Ridge regression under dense factor augmented models.
\newblock {\em Journal of the American Statistical Association},
  119(546):1566--1578.

\bibitem[Hoerl and Kennard, 1970]{hoerl1970ridge}
Hoerl, A.~E. and Kennard, R.~W. (1970).
\newblock Ridge regression: Biased estimation for nonorthogonal problems.
\newblock {\em Technometrics}, 12(1):55--67.

\bibitem[Johnson and Lindenstrauss, 1984]{johnson1984extensions}
Johnson, W.~B. and Lindenstrauss, J. (1984).
\newblock Extensions of {L}ipschitz mappings into a {H}ilbert space.
\newblock {\em Contemporary Mathematics}, 26(189-206):1.

\bibitem[Johnstone and Lu, 2009]{johnstone2009consistency}
Johnstone, I.~M. and Lu, A.~Y. (2009).
\newblock On consistency and sparsity for principal components analysis in high
  dimensions.
\newblock {\em Journal of the American Statistical Association},
  104(486):682--693.

\bibitem[Karabiyik and Westerlund, 2021]{karabiyik2020forecasting}
Karabiyik, H. and Westerlund, J. (2021).
\newblock Forecasting using cross-section average--augmented time series
  regressions.
\newblock {\em Econometrics Journal}, 24(2):313--333.

\bibitem[Koop et~al., 2019]{koop2016bayesian}
Koop, G., Korobilis, D., and Pettenuzzo, D. (2019).
\newblock Bayesian compressed vector autoregressions.
\newblock {\em Journal of Econometrics}, 210(1):135--154.

\bibitem[Liu et~al., 2023]{liu2023random}
Liu, C., Zhao, X., and Huang, J. (2023).
\newblock A random projection approach to hypothesis tests in high-dimensional
  single-index models.
\newblock {\em Journal of the American Statistical Association},
  119(546):1008--1018.

\bibitem[McCracken and Ng, 2020]{mccracken2020fred}
McCracken, M. and Ng, S. (2020).
\newblock {FRED-QD}: {A} quarterly database for macroeconomic research.
\newblock Working Paper 26872, National Bureau of Economic Research.

\bibitem[McCracken and Ng, 2016]{mccracken2015fred}
McCracken, M.~W. and Ng, S. (2016).
\newblock {FRED}-{MD}: A monthly database for macroeconomic research.
\newblock {\em Journal of Business \& Economic Statistics}, 34(4):574--589.

\bibitem[Onatski, 2010]{onatski2010determining}
Onatski, A. (2010).
\newblock Determining the number of factors from empirical distribution of
  eigenvalues.
\newblock {\em Review of Economics and Statistics}, 92(4):1004--1016.

\bibitem[Onatski, 2012]{onatski2012asymptotics}
Onatski, A. (2012).
\newblock Asymptotics of the principal components estimator of large factor
  models with weakly influential factors.
\newblock {\em Journal of Econometrics}, 168(2):244--258.

\bibitem[Paul, 2007]{paul2007asymptotics}
Paul, D. (2007).
\newblock Asymptotics of sample eigenstructure for a large dimensional spiked
  covariance model.
\newblock {\em Statistica Sinica}, 17(4):1617--1642.

\bibitem[Schneider and Gupta, 2016]{schneider2016forecasting}
Schneider, M.~J. and Gupta, S. (2016).
\newblock Forecasting sales of new and existing products using consumer
  reviews: A random projections approach.
\newblock {\em International Journal of Forecasting}, 32(2):243--256.

\bibitem[Stock and Watson, 1998]{stock1998diffusion}
Stock, J.~H. and Watson, M.~W. (1998).
\newblock Diffusion indexes.
\newblock Working Paper 6702, National Bureau of Economic Research.

\bibitem[Stock and Watson, 2002a]{stock2002forecasting}
Stock, J.~H. and Watson, M.~W. (2002a).
\newblock Forecasting using principal components from a large number of
  predictors.
\newblock {\em Journal of the American Statistical Association},
  97(460):1167--1179.

\bibitem[Stock and Watson, 2002b]{stock2002macroeconomic}
Stock, J.~H. and Watson, M.~W. (2002b).
\newblock Macroeconomic forecasting using diffusion indexes.
\newblock {\em Journal of Business \& Economic Statistics}, 20(2):147--162.

\bibitem[Uematsu and Yamagata, 2022a]{uematsu2022estimation}
Uematsu, Y. and Yamagata, T. (2022a).
\newblock Estimation of sparsity-induced weak factor models.
\newblock {\em Journal of Business \& Economic Statistics}, 41(1):213--227.

\bibitem[Uematsu and Yamagata, 2022b]{uematsu2022inference}
Uematsu, Y. and Yamagata, T. (2022b).
\newblock Inference in sparsity-induced weak factor models.
\newblock {\em Journal of Business \& Economic Statistics}, 41(1):126--139.

\end{thebibliography}


\ifthenelse{\boolean{APPON}}{
    \newpage
    \clearpage
    \newpage
\begin{appendices}
\crefalias{section}{appsec}
\crefalias{subsection}{appsubsec}
\numberwithin{equation}{section}

\section{Proofs}
\label{AppendixProofs}

\subsection{Preliminary results: some matrix algebra}
The following identity appears for example in \citet{brinkhuis2005matrix}. For completeness, we provide a proof below.
\begin{lemma}\label{lem:brinkhuis}
    \begin{equation}
        \begin{split}
            &(t\bs Y_{1}+(1-t)\bs Y_{2})^{-1} =t\bs Y_{1}^{-1}+(1-t)\bs Y_{2}^{-1} \\
            &\quad -t(1-t)(\bs Y_{1}^{-1}-\bs Y_{2}^{-1})(t\bs Y_{2}^{-1}+(1-t)\bs Y_{1}^{-1})^{-1}(\bs Y_{1}^{-1}-\bs Y_{2}^{-1}).
        \end{split}
    \end{equation}
\end{lemma}
\textit{Proof:} First rewrite
\begin{equation}
    \begin{split}
        t\bs Y_{1}+(1-t)\bs Y_{2}& = \bs Y_{1}(\bs Y_{1}^{-1}+t(\bs Y_{2}^{-1}-\bs Y_{1}^{-1})\bs Y_{2})\\
        &=\bs Y_{1}(t\bs Y_{2}^{-1}+(1-t)\bs Y_{1}^{-1})\bs Y_{2}.
    \end{split}
\end{equation}
Then,
\begin{equation}
    \begin{split}
        &(t\bs Y_{1}+(1-t)\bs Y_{2})^{-1} \\
        &= \bs Y_{2}^{-1}(t\bs Y_{2}^{-1}+(1-t)\bs Y_{1}^{-1})^{-1}\bs Y_{1}^{-1}\\
        &=\bs Y_{2}^{-1}-\bs Y_{2}^{-1} + \bs Y_{2}^{-1}(t\bs Y_{2}^{-1}+(1-t)\bs Y_{1}^{-1})^{-1}\bs Y_{1}^{-1}\\
        & = \bs Y_{2}^{-1}-\bs Y_{2}^{-1} (t\bs Y_{2}^{-1}+(1-t)\bs Y_{1}^{-1})^{-1}(t\bs Y_{2}^{-1}+(1-t)\bs Y_{1}^{-1}) + \bs Y_{2}^{-1}(t\bs Y_{2}^{-1}+(1-t)\bs Y_{1}^{-1})^{-1}\bs Y_{1}^{-1}\\
        &  =\bs Y_{2}^{-1}-t\bs Y_{2}^{-1}(t\bs Y_{2}^{-1}+(1-t)\bs Y_{1}^{-1})^{-1}(\bs Y_{2}^{-1}-\bs Y_{1}^{-1})\\
        &=\bs Y_{2}^{-1}+t(\bs Y_{1}^{-1}-\bs Y_{2}^{-1})-t(\bs Y_{1}^{-1}-\bs Y_{2}^{-1}) -t\bs Y_{2}^{-1}(t\bs Y_{2}^{-1}+(1-t)\bs Y_{1}^{-1})^{-1}(\bs Y_{2}^{-1}-\bs Y_{1}^{-1})\\
        & = \bs Y_{2}^{-1}+t(\bs Y_{1}^{-1}-\bs Y_{2}^{-1})-t(t\bs Y_{2}^{-1}+(1-t)\bs Y_{1}^{-1}-\bs Y_{2}^{-1})(t\bs Y_{2}^{-1}+(1-t)\bs Y_{1}^{-1})^{-1}(\bs Y_{1}^{-1}-\bs Y_{2}^{-1})\\
        & =t\bs Y_{1}^{-1}+(1-t)\bs Y_{2}^{-1} -t(1-t)(\bs Y_{1}^{-1}-\bs Y_{2}^{-1})(t\bs Y_{2}^{-1}+(1-t)\bs Y_{1}^{-1})^{-1}(\bs Y_{1}^{-1}-\bs Y_{2}^{-1}).
    \end{split}
\end{equation}
This completes the proof.\hfill$\blacksquare$

\begin{lemma}\label{res:matrixconvex}
    Consider a symmetric positive definite matrix $\bs Y$ and a fixed vector $\bs x$. The function
    \begin{equation}
        f(\bs Y) = \frac{\bs x'\bs Y^{-1}\bs x}{1+\bs x'\bs Y^{-1}\bs x}
    \end{equation}
    is matrix convex in $\bs Y$.
\end{lemma}
\textit{Proof.} For matrix convexity, we need to show that for all $0\leq t\leq 1$,
\begin{equation}\label{eq:matconv}
    \frac{\bs x'(t\bs Y_{1}+(1-t)\bs Y_{2})^{-1}\bs x}{1+\bs x'(t\bs Y_{1}+(1-t)\bs Y_{2})^{-1}\bs x}\leq t\frac{\bs x'\bs Y_{1}^{-1}\bs x}{1+\bs x'\bs Y_{1}^{-1}\bs x} + (1-t)\frac{\bs x'\bs Y_{2}^{-1}\bs x}{1+\bs x'\bs Y_{2}^{-1}\bs x}.
\end{equation}
For notational convenience, define $\bs A = (t\bs Y_{1}+(1-t)\bs Y_{2})^{-1}$ for the moment. Then, \eqref{eq:matconv} is equivalent to
\begin{equation}
    \begin{split}
        & \bs x'\bs A\bs x(1+\bs x'\bs Y_{1}^{-1}\bs x)(1+\bs x'\bs Y_{2}^{-1}\bs x)\\
        &\leq t\bs x'\bs Y_{1}^{-1}\bs x(1+\bs x'\bs Y_{2}^{-1}\bs x)+(1-t)\bs x'\bs Y_{2}^{-1}\bs x(1+\bs x'\bs Y_{1}^{-1}\bs x)\\
        &\quad +\bs x'\bs A\bs x\left[t\bs x'\bs Y_{1}^{-1}\bs x(1+\bs x'\bs Y_{2}^{-1}\bs x) + (1-t)\bs x'\bs Y_{2}^{-1}\bs x(1+\bs x'\bs Y_{1}^{-1}\bs x)\right].
    \end{split}
\end{equation}
Gathering the terms with $\bs x'\bs A\bs x$ and simplifying, we get
\begin{equation}
    \begin{split}
        \bs x'\bs A\bs x(1+(1-t)\bs x'\bs Y_{1}^{-1}\bs x + t\bs x'\bs Y_{2}^{-1}   \bs x)\leq t\bs x'\bs Y_{1}^{-1}\bs x + (1-t)\bs x'\bs Y_{2}^{-1}\bs x +\bs x'\bs Y_{1}^{-1}\bs x\bs x'\bs Y_{2}^{-1}\bs x.
    \end{split}
\end{equation}
Define $\bs B = (\bs Y_{1}^{-1}-\bs Y_{2}^{-1})(t\bs Y_{2}^{-1}+(1-t)\bs Y_{1}^{-1})^{-1}(\bs Y_{1}^{-1}-\bs Y_{2}^{-1}).$ Now using \cref{lem:brinkhuis}, we get
\begin{equation}
    \begin{split}
        &-t(1-t)\bs x'\bs B\bs x + (1-t)t(\bs x'\bs Y_{1}^{-1}\bs x)^2 + (1-t)^2(\bs x'\bs Y_{1}^{-1}\bs x)(\bs x'\bs Y_{2}^{-1}\bs x)\\
        & -t(1-t)^2\bs x'\bs B\bs x\bs x'\bs Y_{1}^{-1}\bs x\\
        &+t^2\bs x'\bs Y_{1}^{-1}\bs x \bs x'\bs Y_{2}^{-1}\bs x + t(1-t)(\bs x'\bs Y_{2}^{-1}\bs x)^2-t^2(1-t)\bs x'\bs B\bs x\bs x'\bs Y_{2}^{-1}\bs x\\
        &\leq \bs x'\bs Y_{1}^{-1}\bs x\bs x'\bs Y_{2}^{-1}\bs x.
    \end{split}
\end{equation}
Rewriting some more, we see that this is equivalent to
\begin{equation}
    \begin{split}
        (\bs x'(\bs Y_{1}^{-1}-\bs Y_{2}^{-1})\bs x)^2\leq \bs x'\bs B\bs x (1+(1-t)\bs x'\bs Y_{1}^{-1}\bs x + t\bs x'\bs Y_{2}^{-1}\bs x).
    \end{split}
\end{equation}
Since $\bs x'\bs B\bs x\geq 0$, it is sufficient to show that
\begin{equation}\label{eq:equiv}
    \begin{split}
        (\bs x'(\bs Y_{1}^{-1}-\bs Y_{2}^{-1})\bs x)^2&\leq \bs x'(\bs Y_{1}^{-1}-\bs Y_{2}^{-1})[(1-t)\bs Y_{1}^{-1}+t\bs Y_{2}^{-1})]^{-1}(\bs Y_{1}^{-1}-\bs Y_{2}^{-1})\bs x\\
        &\qquad \cdot \bs x'[(1-t)\bs Y_{1}^{-1}+t\bs Y_{2}^{-1})]\bs x.
    \end{split}
\end{equation}
However, this inequality follows immediately from the Cauchy-Schwarz inequality, as we can write the left-hand side as
\begin{equation}
    \begin{split}
        &(\bs x'(\bs Y_{1}^{-1}-\bs Y_{2}^{-1})\bs x)^2 = (\bs x'(\bs Y_{1}^{-1}-\bs Y_{2}^{-1})[(1-t)\bs Y_{1}^{-1}+t\bs Y_{2}^{-1})]^{-1/2}[(1-t)\bs Y_{1}^{-1}+t\bs Y_{2}^{-1})]^{1/2}\bs x)^2.
    \end{split}
\end{equation}
This completes the proof.
\hfill$\blacksquare$

\subsection{Preliminary results: expectations over random projections}
\begin{lemma}\label{LemmaExpectationSwitch} Let $\bs R\in \mathbb{R}^{N\times k}$ be a matrix with independent entries $R_{ij}\sim N(0,1)$. Consider a matrix $\bs Z\in\mathbb{R}^{T\times N}$ with singular value decomposition $\bs Z = \bs U\bs D\bs V'$. Then,
    \begin{equation}
        \mathbb{E}_R[\bs R(\bs R'\bs Z'\bs Z\bs R)^{-1}\bs R'] = \bs V \mathbb{E}_R[\bs R(\bs R'\bs D'\bs D \bs R)^{-1}\bs R']\bs V'.
    \end{equation}
\end{lemma}
\textit{Proof.} Define $\tilde{\bs R} = \bs V'\bs R$. Using that $\bs V\bs V'=\bs I$, we have that,
\begin{equation}
    \bs R(\bs R'\bs V\bs D'\bs D\bs V'\bs R)^{-1}\bs R'=\bs V\tilde{\bs R}(\tilde{\bs R}'\bs D'\bs D\tilde{\bs {R}})^{-1}\tilde{\bs R}'\bs V'.
\end{equation}
Since the elements of $\bs R$ are i.i.d.\ standard normal random variables, so are the elements of $\tilde{\bs R}$. Redefining $\tilde{\bs R}$ as $\bs R$ gives the result.\hfill$\blacksquare$

\begin{lemma}\label{ProofOffDiag0} Consider the notation as in \cref{LemmaExpectationSwitch} and let $\bs b_{i}\in \mathbb{R}^{N}$ be a vector with value 1 at the $i$th position and zeros elsewhere. Then, $
    \bs b_{i}'\mathbb{E}_R[\bs R(\bs R'\bs D'\bs D\bs R)^{-1}\bs R']\bs b_{j}=0$ for $i\neq j$.
\end{lemma}
\textit{Proof.} Since the elements of $\bs R$ are i.i.d. standard normal, $\bs R\overset{(d)}{=}\bs Z\bs R$, where $\bs Z$ is a diagonal matrix that is independent of $\bs R$ and with independent Rademacher random variables on the diagonal. Since $\bs D'\bs D$ is diagonal,
\begin{equation}
    \begin{split}
        \bs b_{i}'\mathbb{E}_R[\bs R(\bs R'\bs D'\bs D\bs R)^{-1}\bs R']\bs b_{j}&= \bs b_{i}'\mathbb{E}_{Z,R}[\bs Z\bs R(\bs R'\bs D'\bs D\bs R)^{-1}\bs R'\bs Z']\bs b_{j}\\
        &= \bs b_{i}'\mathbb{E}_{Z}[z_{i}z_{j}]\mathbb{E}_{R}[\bs R(\bs R'\bs D'\bs D\bs R)^{-1}\bs R']\bs b_{j}=0.
    \end{split}
\end{equation}
\hfill $\blacksquare$

\begin{lemma}\label{lem:bounddiagonal} Let $\bs D$ be an $l\times l$ diagonal matrix with elements $[\bs D^2]_{ii}=d_{i}$ such that $d_{i}\geq d_{i+1}$ and $d_{i}\geq C>0$ for $i=1,\ldots,m$. Let $\bs R$ a $l\times k$ matrix with i.i.d.\ standard normal entries. Assume that $k<m-2$. Consider
    \begin{equation}
        w_{i} = \bs e_{i}'\bs D\mathbb{E}_R[\bs R(\bs R'\bs D^2\bs R)^{-1}\bs R']\bs D\bs e_{i},
    \end{equation}
    We can bound $w_{i}$ for $i=1,\ldots,m$ as
    \begin{equation}
        1-\frac{1}{1+(k-2)\frac{d_{i}}{\sum_{j\neq i}d_{j}}}\leq w_{i} \leq  1-\frac{1}{1+(d_{i}/d_{m})k/(m-k-2)}
    \end{equation}
\end{lemma}
\textit{Proof.} We first show the first inequality. Denote $\bs{R} = [\bs{r}_1, \bs{r}_2,..., \bs{r}_l ]'$ where $\bs{r}_i$ are $k \times 1$ vectors. Define $\bs M=\bs R'\bs D^2 \bs R$. Then,
\begin{equation}
    \bs{M} =d_i \bs{r}_i \bs{r}_i'+ \sum_{j\neq i}d_{j} \bs{r}_j \bs{r}_j'  =d_i \bs{r}_i \bs{r}_i'+ \bs{M}_{-i}.
\end{equation}
By applying the Sherman-Morrison formula and using \cref{res:matrixconvex} and the fact that for any matrix convex function we have $\mathbb{E}[h(\bs Z)]\geq h(\mathbb{E}[\bs Z])$, we see that
\begin{equation}
    \label{eq: SMF to offdiag}
    \begin{split}
        \mathbb{E}_{R}[d_{i}\bs{r}_i'\bs{M}^{-1}\bs{r}_i ] &= \mathbb{E}_{R} \left[ \frac{ d_{i}\bs{r}_i'  \bs{M}_{-i}^{-1} \bs{r}_i }{1 + d_{i} \bs{r}_i'  \bs{M}_{-i}^{-1} \bs{r}_i } \right]\\
        &\geq \mathbb{E}_{r_{i}}\left[\frac{d_{i}\bs r_{i}'\mathbb{E}_{R_{-i}}[\bs M_{-i}]^{-1}\bs r_{i}}{1+d_{i}\bs r_{i}'\mathbb{E}_{R_{-i}}[\bs M_{-i}]^{-1}\bs r_{i}}\right].\\
        & = \mathbb{E}_{r_{i}}\left[\frac{\frac{d_{i}}{\sum_{j\neq i}d_{j}}\bs r_{i}'\bs r_{i}}{1+\frac{d_{i}}{\sum_{j\neq i}d_{j}}\bs r_{i}'\bs r_{i}}\right]\\
        & = 1-\mathbb{E}_{r_{i}}\left[\frac{1}{1+\frac{d_{i}}{\sum_{j\neq i}d_{j}}\bs r_{i}'\bs r_{i}}\right]\\
        &\geq 1-\frac{\mathbb{E}_{r_{i}}[(\bs r_{i}'\bs r_{i})^{-1}]}{\mathbb{E}_{r_{i}}[(\bs r_{i}'\bs r_{i})^{-1}] + \frac{d_{i}}{\sum_{j\neq i}d_{j}}}\\
        &= 1-\frac{1}{1+(k-2)\frac{d_{i}}{\sum_{j\neq i}d_{j}}}.
    \end{split}
\end{equation}

For the second inequality in \cref{lem:bounddiagonal}, first notice that $\bs M\succeq d_m \sum_{j=1}^{m}\bs r_{j}\bs r_{j}'\equiv \tilde{\bs M}$. This implies that $\bs M^{-1}\preceq  \tilde{\bs M}^{-1}$. Then,
\begin{equation}
    \begin{split}
        \mathbb{E}_{R}[d_{i}\bs r_{i}'\bs M^{-1}\bs r_{i}] &\leq \mathbb{E}_{R}[d_{i}\bs r_{i}'\tilde{\bs M}^{-1}\bs r_{i}]\\
        & = \mathbb{E}_{R}\left[\frac{d_{i}\bs r_{i}\tilde{\bs M}^{-1}_{-i}\bs r_{i}}{1+d_{i}\bs r_{i}\tilde{\bs M}^{-1}_{-i}\bs r_{i}}\right]\\
        &\leq \frac{d_{i}\mathbb{E}_{R}\left[\text{tr}\left(\tilde{\bs M}_{-i}^{-1}\right)\right]}{1+d_{i}\mathbb{E}_{R}\left[\text{tr}\left(\tilde{\bs M}_{-i}^{-1}\right)\right]}\\
        &\leq \frac{(d_{i}/d_{m})k/(m-k-2)}{1+(d_{i}/d_{m})k/(m-k-2)},
    \end{split}
\end{equation}
where the last inequality uses the trace of the expectation of a $k$-dimensional inverse Wishart matrix with $m-1$ degrees of freedom.\hfill$\blacksquare$

    \subsection{Preliminary results: factors and loadings}
    We have the following result quantifying the relation between the true factors and the left singular vectors of $\bs Z$ associated with eigenvalues $\lambda_{r+1},\lambda_{r+2},\dots$, as well as the true loadings and the right singular vectors of $\bs Z$ associated with eigenvalues $\lambda_{r+1},\lambda_{r+2},\dots$
    \begin{lemma} Denote $\bs F$, $\tilde{\bs F}$, $\bs U_{-r}$ as in the main text. Let $\bs H_{2}=(\bs F'\bs F)^{-1}\bs F'\tilde{\bs F}$ and $\bs H_{3}=(\tilde{\bs F}'\bs F/T)^{-1}$. Then,
        \begin{align}
            (\bs F'\bs F)^{-1}\bs F'\tilde{\bs F}\tilde{\bs F}'\bs F/T &=  \bs I +O_{p}\left(\frac{1}{\sqrt{N^{\alpha}T}}\right) + O_{p}\left(\frac{1}{N^{\alpha}}\right) + O_{p}\left(\frac{N}{N^{\alpha}}\frac{1}{T}\right),\label{res:BaiNg1}\\
            \|(\bs F'\bs F)^{-1/2}\bs F'\bs U_{-r}\|^2& = O_{p}\left(\frac{1}{\sqrt{N^{\alpha}T}}\right) + O_{p}\left(\frac{1}{N^{\alpha}}\right) + O_{p}\left(\frac{N}{N^{\alpha}}\frac{1}{T}\right),\label{res:helpF}\\
            \|(\bs \Lambda'\bs\Lambda)^{-1}\bs\Lambda'\bs V_{-r}\bs V_{-r}'\bs \Lambda\|&= O_{p}\left(\frac{1}{\sqrt{N^{\alpha}T}}\right) + O_{p}\left(\frac{1}{N^{\alpha}}\right) + O_{p}\left(\frac{N}{N^{\alpha}}\frac{1}{T}\right). \label{res:helpL}
        \end{align}
    \end{lemma}
    \textit{Proof.} The first part is Lemma 3 of \citet{bai2023approximate}. For \eqref{res:helpF}, using the fact that $\bs U\bs U' = \bs I_{T}$, we have
    \begin{align*}
        \bs F'\bs F &= \bs F'\bs U \bs U '\bs F\\
        & = \bs F'\bs U_{r}\bs U_{r}'\bs F + \bs F'\bs U_{-r} \bs U_{-r} '\bs  F\\
        & = \bs F'\tilde{\bs F}\tilde{\bs F}'\bs F/T + \bs F'\bs U_{-r} \bs U_{-r} ' \bs F\\
        & = (\bs F'\bs F)\bs H_{2}\bs H_{3}^{-1} +\bs F'\bs U_{-r} \bs U_{-r} ' \bs F.
    \end{align*}
    Multiplying from the left by $(\bs F'\bs F)^{-1}$ and using Lemma 3 of \citet{bai2023approximate}, we obtain
    \begin{align*}
        \|(\bs F'\bs F)^{-1}\bs F'\bs U_{-r} \bs U_{-r} ' \bs F\| &=\|\bs I-\bs H_{2}\bs H_{3}^{-1}\| \\
        &=O_{p}\left(\frac{1}{\sqrt{N^{\alpha}T}}\right) + O_{p}\left(\frac{1}{N^{\alpha}}\right) + O_{p}\left(\frac{N}{N^{\alpha}}\frac{1}{T}\right).
    \end{align*}
    The proof for \eqref{res:helpL} is almost identical. Using that $\bs V \bs V ' =\bs I_{N}$, we have
    \begin{align*}
        \bs\Lambda'\bs \Lambda &= \bs\Lambda'\bs V \bs V '\bs \Lambda\\
        & =  \bs\Lambda'\bs V_{r} \bs V_{r} '\bs \Lambda+\bs\Lambda'\bs V_{-r} \bs V_{-r} '\bs \Lambda\\
        & = \bs\Lambda'\tilde{\bs \Lambda}\bs D_{r}^{-2}\tilde{\bs \Lambda}'\bs \Lambda /N +\bs\Lambda'\bs V_{-r} \bs V_{-r} '\bs \Lambda\\
        & = \bs H_{4}\bs H_{1}^{-1}\bs\Lambda'\bs \Lambda  + \bs\Lambda'\bs V_{-r} \bs V_{-r} '\bs \Lambda,
    \end{align*}
    where $\bs H_{1}=(\bs\Lambda'\bs\Lambda)(\tilde{\bs\Lambda}'\bs\Lambda)^{-1}$, and $\bs H_{4}=(\bs\Lambda'\tilde{\bs\Lambda}/N)\bs D_{r}^{-2}$. Multiplying from the right by $(\bs \Lambda'\bs \Lambda)^{-1}$ and rearranging, we obtain
    \begin{align*}
        \|(\bs \Lambda'\bs\Lambda)^{-1}\bs\Lambda'\bs V_{-r} \bs V_{-r} '\bs \Lambda\| &= \|\bs I-\bs H_{4}\bs H_{1}^{-1}\| \\
        &=O_{p}\left(\frac{1}{\sqrt{N^{\alpha}T}}\right) + O_{p}\left(\frac{1}{N^{\alpha}}\right) + O_{p}\left(\frac{N}{N^{\alpha}}\frac{1}{T}\right),
    \end{align*}
    where the last equality again follows from Lemma 3 of \citet{bai2023approximate}. \hfill$\blacksquare$

    \begin{lemma}
        \begin{align}
            \frac{1}{N^{\alpha/2}}\|(\bs F'\bs F)^{-1}\bs F'\bs e\bs V_{r} \|&=O_{p}\left(\frac{1}{\sqrt{N^{\alpha} T}}\right)+O_{p}\left(\frac{1}{N^{\alpha}}\right)+O_{p}\left(\frac{N}{N^{\alpha}}\frac{1}{T}\right),\label{res:VNTr}\\
            \frac{1}{N^{\alpha/2}}\|\bs e_{t}'\bs V_{r} \|  &=O_{p}\left(\sqrt{\frac{N}{N^{\alpha}}}\left[\frac{N}{N^{\alpha}}\frac{1}{T} + \frac{1}{N^{\alpha}}+\frac{1}{\sqrt{N}}\right]\right).\label{res:etVNTr}
        \end{align}
    \end{lemma}
    \textit{Proof.} We first prove \eqref{res:VNTr}. Using that $\tilde{\bs\Lambda}=\sqrt{N}\bs V_{r} \bs D_{r}$, $\sqrt{\frac{N}{N^{\alpha}}}\bs D_{r}=O_{p}(1)$, and the fact that $\tilde{\bs\Lambda}=\frac{1}{T}\bs X'\tilde{\bs F}$, we have
    \begin{equation}
        \begin{split}
            \frac{1}{N^{\alpha/2}}\|(\bs F'\bs F)^{-1}\bs F'\bs e\bs V_{r} \|
            &= O_{p}\left(\frac{1}{N^{\alpha}}\right)\|(\bs F'\bs F)^{-1}\bs F'\bs e\tilde{\bs\Lambda}\|\\
            &=\frac{1}{TN^{\alpha}}\|(\bs F'\bs F)^{-1}\bs F'\bs e\bs e'\tilde{\bs F}\| + \frac{1}{TN^{\alpha}}\|(\bs F'\bs F)^{-1}\bs F'\bs e\bs \Lambda\bs F'\tilde{\bs F}\|\\
            &\leq \|(\bs F'\bs F/T)^{-1}\|\|\tilde{\bs F}/\sqrt{T}\|\|\bs F/\sqrt{T}\|\frac{1}{TN^{\alpha}}\|\bs e\bs e'\| \\
            &\quad +\|(\bs F'\bs F/T)^{-1}\|\frac{1}{N^{\alpha}T}\|\bs F'\bs e\bs \Lambda\|\|\bs F'\tilde{\bs F}/T\|\\
            &=O_{p}\left(\frac{1}{N^{\alpha}}\right)+O_{p}\left(\frac{N}{N^{\alpha}}\frac{1}{T}\right)+ O_{p}\left(\frac{1}{\sqrt{N^{\alpha} T}}\right).
        \end{split}
    \end{equation}
    using that $\|\bs e\bs e'\| = O_{p}(\max(N,T))$ by \cref{assA1}, and $\|\bs F'\bs e\bs\Lambda\|=O_{p}(\sqrt{N^{\alpha}T})$ by \cref{assA2}(iii).

    We now consider \eqref{res:etVNTr}.
    \begin{equation}
        \begin{split}
            \frac{1}{N^{\alpha/2}}\|\bs e_{t}'\bs V_{r} \|&=O_{p}\left(\frac{1}{N^{\alpha}}\right)\|\bs e_{t}'\tilde{\bs\Lambda}\|\\
            &=\frac{1}{TN^{\alpha}}\|\bs e_{t}'\bs e'\tilde{\bs F}\| + \frac{1}{TN^{\alpha}}\|\bs e_{t}'\bs \Lambda\bs F'\tilde{\bs F}\|\\
            &=\frac{1}{TN^{\alpha}}\|\bs e_{t}'\bs e'\bs F\bs H_{0}\| + \frac{1}{TN^{\alpha}}\|\bs e_{t}'\bs e'\|\|\tilde{\bs F}-\bs F\bs H_{0}\| + \frac{1}{N^{\alpha}}\|\bs e_{t}'\bs\Lambda\|\\
            &=O_{p}\left(\frac{N}{N^{\alpha}}\delta_{NT}^{-2}\right)+O_{p}\left(\frac{N}{N^{\alpha}}\delta_{NT}^{-1}\left(\frac{1}{N^{\alpha/2}}+\frac{1}{T}\frac{N}{N^{\alpha}}\right)\right) +O_{p}\left(\frac{1}{\sqrt{N^{\alpha}}}\right).\\
        \end{split}
    \end{equation}
    where the first term and final term follow from \cref{corr1} (ii), and the second term follows from \cref{corr1} (i) and Proposition 1 in \citet{bai2023approximate}.

    \begin{lemma}
        \begin{equation}
            \frac{1}{\sqrt{T}}\|\bs U_{r}\bs\varepsilon\| =O_{p}\left(\frac{1}{T^{1/2}}\right).
        \end{equation}
    \end{lemma}
    \textit{Proof}.
    \begin{equation}
        \begin{split}
            \frac{1}{\sqrt{T}}\|\bs U_{r}\bs\varepsilon\|& = \frac{1}{T}\|\tilde{\bs F}'\bs\varepsilon\|=O_{p}\left(\frac{1}{T^{1/2}}\right).
        \end{split}
    \end{equation}
    where last line uses \cref{ass:varepsilon} and Proposition 1 in \citet{bai2023approximate}.
    \begin{lemma} Under \cref{assA1}--\cref{assA5},
 \begin{equation} 
        \sqrt{N^{\alpha}}(\tilde{\bs f}_{T}-\bs H_{4}'\bs f_{T}) = O_{p}(1) + O_{p}\left(\sqrt{\frac{N}{N^{\alpha}}\frac{1}{T}}\right) + O_{p}\left(\frac{1}{N^{\alpha}}\right).
    \end{equation}
    \end{lemma}
   \textit{Proof}: \citet{bai2023approximate} Proposition 2 shows
    \begin{align}
        \frac{1}{N}\|\tilde{\bs\Lambda}-\bs\Lambda(\bs H_{0}')^{-1}\|^2 &= O_{p}\left(\frac{1}{T}\right)+ O_{p}\left(\frac{1}{N^{1+\alpha}}\right),\\
        \frac{1}{N^{\alpha}}\|\tilde{\bs\Lambda}-\bs\Lambda(\bs H_{0}')^{-1}\|^2 &= O_{p}\left(\frac{N}{N^{\alpha}}\frac{1}{T}\right)+ O_{p}\left(\frac{1}{N^{2\alpha}}\right).
    \end{align}
    Under \cref{assA5}, we then have
    \begin{equation}
        \frac{1}{N^{\alpha/2}}\|(\tilde{\bs\Lambda}-\bs\Lambda(\bs H_{4}')^{-1})'\bs e_{T}\| = O_{p}\left(\sqrt{\frac{N}{N^{\alpha}}\frac{1}{T}}\right)+ O_{p}\left(\frac{1}{N^{\alpha}}\right).
    \end{equation}
    this implies the result.
   
    \subsection{Preliminary results: random projection regression}
    \begin{lemma}\label{res:D}
        \begin{enumerate}
            \item  Define $\bs M=\bs X\mathbb{E}_{R}[\bs R(\bs R'\bs X'\bs X\bs R)^{-1}\bs R']\bs X'$. Then, $\bs M=\bs U \bs D\bs U '$ where $[\bs D]_{ii}=1+O_{p}\left(\frac{N}{N^{\alpha}}k^{-1}\right)$ for $i=1,\ldots,r$ and $[\bs D]_{ii} = O_{p}(k\delta_{NT}^{-2})$ for $i=r+1,\ldots,\delta_{NT}^2$.
            \item Define $\bs D_{2} = \mathbb{E}[\bs R(\bs R'\bs D_{NT}'\bs D_{NT}\bs R)^{-1}\bs R']\bs D_{NT}'$.
            Then,  $[\bs D_{2}]_{ii}=O_{p}\left(\sqrt{\frac{N}{N^{\alpha}}}\right)$ for $i=1,\ldots,r$ and $[\bs D_{2}]_{ii}=O_{p}(k\delta_{NT}^{-1})$ for $i=r+1,\ldots,\delta_{NT}^2$.
        \end{enumerate}
    \end{lemma}
    \textit{Proof.}
    We only prove part (i), part (ii) follows analogously. Recall that $\bs X/\sqrt{NT}=\bs U \bs D_{NT}\bs V '$ by \eqref{eq:defZ}. Using \cref{LemmaExpectationSwitch}.
    \begin{equation}
        \begin{split}
            \bs M&=\bs X\mathbb{E}_{R}[\bs R(\bs R'\bs X'\bs X\bs R)^{-1}\bs R']\bs X'\\
            &=\bs U \bs D_{NT}\mathbb{E}_{R}[\bs R(\bs R'\bs D_{NT}'\bs D_{NT}\bs R)^{-1}\bs R']\bs D_{NT}'\bs U '\\
            &=\bs U \bs D\bs U '.
        \end{split}
    \end{equation}
    By \cref{lem:bounddiagonal}, the fact that the first $r$ eigenvalues of $\bs X'\bs X/NT$ are $O_{p}(N^{\alpha}/N)$ and the next $\delta_{NT}^2-r$ eigenvalues are $O_{p}(\delta_{NT}^{-2})$ as established in \citet[Lemma A.9]{ahn2013eigenvalue}, $\bs D$ is a diagonal matrix with $[\bs D]_{ii} = 1+O_{p}\left(\frac{N}{N^{\alpha}}k^{-1}\right)$ for $i=1,\ldots,r$ and  $[\bs D]_{ii}=O_{p}(k\cdot \delta_{NT}^{-2})$ for $i=r+1,\ldots,\delta_{NT}^2$.

    \section{Proofs of the main results}
    \subsection{Proof of Theorem 1 and Theorem 3}
    We have
    \begin{equation}\label{eq:proofthm3}
        \begin{split}
            \hat{y}_{T+h|T}^{pca}& = \tilde{\bs f}_{T}('\tilde{\bs F}'\tilde{\bs F})^{-1}\tilde{\bs F}'\bs y
            \\
            & = \tilde{\bs f}_{T}'(\tilde{\bs F}'\tilde{\bs F})^{-1}\tilde{\bs F}'\bs F\bs \gamma + \tilde{\bs f}_{T}'(\tilde{\bs F}'\tilde{\bs F})^{-1}\tilde{\bs F}'\bs \varepsilon\\
            & = \bs f_{T}'\bs H_{0}(\tilde{\bs F}'\tilde{\bs F})^{-1}\tilde{\bs F}'`\bs F\bs \gamma +  (\tilde{\bs f}_{T}'-\bs f_{T}'\bs H_{0})(\tilde{\bs F}'\tilde{\bs F})^{-1}\tilde{\bs F}'\bs F\bs \gamma + \tilde{\bs f}_{T}'(\tilde{\bs F}'\tilde{\bs F})^{-1}\tilde{\bs F}'\bs \varepsilon\\
            &=\bs f_{T}'\bs H_{0}\bs H_{3}^{-1}\bs \gamma +  (\tilde{\bs f}_{T}'-\bs f_{T}'\bs H_{0})(\tilde{\bs F}'\tilde{\bs F})^{-1}\tilde{\bs F}'\bs F\bs \gamma + \tilde{\bs f}_{T}'(\tilde{\bs F}'\tilde{\bs F})^{-1}\tilde{\bs F}'\bs \varepsilon\\
            &=\bs f_{T}'\bs \gamma + O_{p}\left(\frac{1}{\sqrt{N^{\alpha}T}}\right) + O_{p}\left(\frac{1}{{N^{\alpha}}}\right)+ O_{p}\left(\frac{N}{{N^{\alpha}T}}\right) + \|\tilde{\bs f}_{T}'-\bs f_{T}'\bs H_{0}\|O_{p}(1) + O_{p}\left(\frac{1}{\sqrt{T}}\right)\\
            &=\bs f_{T}'\bs \gamma+ O_{p}\left(\frac{1}{N^{\alpha/2}}\right)+O_{p}\left(\frac{N^{3/2}}{N^{3\alpha/2}}\frac{1}{T}\right)+ O_{p}\left(\frac{N^{1/2}}{N^{3\alpha/2}}\right)+O_{p}\left(\frac{N^2}{N^{2\alpha}T^{1/2}}\right)\\
            &\quad +O_{p}\left(\frac{N^2}{N^{2\alpha}T^{3/2}}\right)+ O_{p}\left(\frac{1}{\sqrt{T}}\right),
        \end{split}
    \end{equation}
    where we use that $\bs H_{0}\bs H_{3}^{-1} = \bs I+O_{p}\left(\frac{1}{\sqrt{N^{\alpha}T}}\right) + O_{p}\left(\frac{1}{{N^{\alpha}}}\right)+ O_{p}\left(\frac{N}{{N^{\alpha}T}}\right)$ by Lemma 3 in \citet{bai2023approximate}, and $\|\tilde{\bs f}_{T}'-\bs f_{T}'\bs H_{0}\|=O_{p}\left(\frac{1}{N^{\alpha/2}}\right)+O_{p}\left(\frac{N^{3/2}}{N^{3\alpha/2}}\frac{1}{T}\right)+ O_{p}\left(\frac{N^{1/2}}{N^{3\alpha/2}}\right)+O_{p}\left(\frac{N^2}{N^{2\alpha}T^{1/2}}\right) +O_{p}\left(\frac{N^2}{N^{2\alpha}T^{3/2}}\right)$. To obtain the latter result, \citet[p. 1901]{bai2023approximate} shows that
    \begin{equation}\label{eqB2}
        \sqrt{N^{\alpha}}(\tilde{\bs f}_{T}'-\bs f_{T}'\bs H_{0})=\left(\frac{\tilde{\bs \Lambda}'\tilde{\bs \Lambda}}{N^{\alpha}}\right)^{-1}\bs H_{4}^{-1}\sum_{i=1}^{N}\bs\Lambda_{i}e_{iT} +\left(\frac{\tilde{\bs \Lambda}'\tilde{\bs \Lambda}}{N^{\alpha}}\right)^{-1}\frac{(\tilde{\bs\Lambda}-\bs\Lambda\bs (\bs H_{4}')^{-1})'\bs e_T}{\sqrt{N^{\alpha}}}.
    \end{equation}
    The term on the right-hand side is $O_{p}(1)$ by Corollary 1 part (ii). The second term is covered by \citet[Lemma A.2]{bai2023approximate}. This yields the result in  \eqref{eq:proofthm3} and completes the proof for \cref{thm:0}.
    
    Under \cref{assA5}, the idiosyncratic errors are independent over time. In that case, we can establish a much faster rate on the second term on the right-hand side of \eqref{eqB2}. From our construction of the PCA-based forecast, $\tilde{\bs\Lambda}$ is estimated on data up to and including time $t-h$, and hence, independent of $\bs e_{t}$. Then,
    \begin{equation}
        \begin{split} \mathbb{E}\left[\left(\frac{(\tilde{\bs\Lambda}-\bs\Lambda\bs (\bs H_{4}')^{-1})'\bs e_T}{\sqrt{N^{\alpha}}}\right)^2\right] &= \frac{1}{N^{\alpha}}\mathbb{E}\left[(\tilde{\bs\Lambda}-\bs\Lambda\bs (\bs H_{4}')^{-1})'\bs G_{N}(\tilde{\bs\Lambda}-\bs\Lambda\bs (\bs H_{4}')^{-1})\right]\\
            &\leq\frac{C}{N^{\alpha}}\mathbb{E}\left[(\tilde{\bs\Lambda}-\bs\Lambda\bs (\bs H_{4}')^{-1})'(\tilde{\bs\Lambda}-\bs\Lambda\bs (\bs H_{4}')^{-1})\right]\\
            &=O\left(\frac{N}{N^\alpha}\frac{1}{T}\right)+ O\left(\frac{1}{N^{2\alpha}}\right).
        \end{split}
    \end{equation}
    Using \cref{assA4}, we then have
    \begin{equation}
        \|\tilde{\bs f}_{T}'-\bs f_{T}'\bs H_{0}\| = O_{p}\left(\frac{1}{N^{\alpha/2}}\right).
    \end{equation}
    This completes the proof of \cref{thm:2}.
    \subsection{Proof of Theorem 2 and Theorem 4}

    Using \eqref{eq:factorequation} and the fact that $
    (\bs F'\bs F)^{-1}\bs F'\bs X = \bs\Lambda' + (\bs F'\bs F)^{-1} \bs F'\bs e$, we can rewrite the forecast \eqref{eq:rpforecast} as
    \begin{equation}\label{eq:rpexpansion}
        \begin{split}
            \hat{y}_{T+h|T}^{rp}    & = \bs f_{T}'\bs\Lambda'\mathbb{E}_{R}[\bs R(\bs R'\bs X'\bs X\bs R)^{-1}\bs R']\bs X'\bs F\bs \gamma + \bs e_{T}'\mathbb{E}_{R}[\bs R(\bs R'\bs X'\bs X\bs R)^{-1}\bs R']\bs X'\bs F\bs \gamma\\
            &\quad +\bs x_{T}'\mathbb{E}_{R}[\bs R(\bs R'\bs X'\bs X\bs R)^{-1}\bs R']\bs X'\bs \varepsilon\\
            & = \bs f_{T}'\bs\gamma + \underbrace{\bs f_{T}'\big[(\bs F'\bs F)^{-1}\bs F'\bs M\bs F-\bs I\big]\bs \gamma}_{(I)}- \underbrace{\bs f_{T}'(\bs F'\bs F)^{-1}\bs F'\bs e\bs M_{2}\bs F\bs \gamma}_{(II)}+\underbrace{\bs e_{T}'\bs M_{2}\bs F\bs \gamma}_{(III)}-\underbrace{\bs b_{T}'\bs M\bs\varepsilon}_{(IV)}.
        \end{split}
    \end{equation}
    where $\bs M=\bs X\mathbb{E}_{R}[\bs R(\bs R'\bs X'\bs X\bs R)^{-1}\bs R']\bs X'=\bs U \bs D\bs U '$. \cref{res:D} shows that $[\bs D]_{ii} = 1+O_{p}\left(\frac{N}{N^{\alpha}}\frac{1}{k}\right)$ for $i=1,\ldots,r$ and $[\bs D]_{ii} =O_{p}\left(k\delta_{NT}^{-2}\right)$ for $i>r$. We also define $$\bs M_{2} =\mathbb{E}_{R}[\bs R(\bs R'\bs X'\bs X\bs R)^{-1}\bs R']\bs X'=\frac{1}{\sqrt{NT}} \bs V \bs D_{2}\bs U ',$$ with $[\bs D_2]_{ii} = O_{p}\left(\frac{N}{N^{\alpha}}\right)$ for $i=1,\ldots,r$ and $[\bs D]_{ii} =O_{p}\left(k\delta_{NT}^{-1}\right)$ for $i>r$. $\bs b_{T}$ is a $T\times 1$ vector with the $T$th element equal to one and the only nonzero element. The expansion \eqref{eq:rpexpansion} also holds for the ridge forecast \eqref{eq:ridgeforecast}.

    We now bound the terms $(I)-(IV)$. The difference between \cref{thm:1} and \cref{thm:3} is in the analysis of  term $(III)$.
    \paragraph{Term $(I)$}
    Assuming that $k\delta_{NT}^{-2}\rightarrow 0$ and $\frac{N}{N^{\alpha}}k^{-1}\rightarrow 0$, 
    \begin{align*}
        \|(I)\|
        & = \|\bs f_{T}'\|\|(\bs F'\bs F)^{-1}\bs F'\bs U \bs D\bs U '\bs F-\bs I\|\|\bs\gamma\|\\
        &\leq \|\bs f_{T}'\|\|(\bs F'\bs F)^{-1}\bs F'\bs U_{r}\bs U_{r}'\bs F-\bs I\|\|\bs\gamma\| +O_{p}\left(\frac{N}{N^{\alpha}}k^{-1}\right) \|\bs f_{T}'\|\|(\bs F'\bs F/T)^{-1}\|\frac{1}{T}\|\bs F'\bs U_{r}\|^2\|\bs\gamma\| \\
        &\quad  + O_{p}(k\cdot \delta_{NT}^{-2})\|\bs f_{T}'\|\|(\bs F'\bs F)^{-1/2}\bs F'\bs U_{-r} \|^2\|\bs \gamma\|\\
        & = O_{p}\left(\frac{1}{\sqrt{N^{\alpha}T}}\right) + O_{p}\left(\frac{1}{N^{\alpha}}\right) + O_{p}\left(\frac{N}{N^{\alpha}}\frac{1}{T}\right) + O_{p}\left(\frac{N}{N^{\alpha}}k^{-1}\right).
    \end{align*}
    To obtain the final line, we applied \cref{res:BaiNg1} to the first and second term after the inequality sign, and we applied \cref{res:helpF} to the third term after the inequality sign.

    \paragraph{Term $(II)$}  We have,
    \begin{align*}
        \|(II)\| & = \|\bs f_{T}'(\bs F'\bs F/T)^{-1}\frac{1}{T}\bs F'\bs e \frac{1}{\sqrt{NT}}\bs V \bs D_{2} \bs U '\bs F\bs \gamma\| \\
        &\leq \|\bs f_{T}\|\|\bs \gamma\|\frac{1}{\sqrt{N^{\alpha}}}\|(\bs F'\bs F)^{-1}\bs F'\bs e\bs V_{r} \|\frac{1}{\sqrt{T}}\|\bs U_{r}\bs F\|O_{p}(1) + \\
        &\quad O_{p}(k\delta_{NT}^{-1})\|\bs f_{T}\|\|\bs \gamma\|\frac{1}{\sqrt{NT}}\|\bs F'\bs e\|\frac{1}{\sqrt{T}}\|(\bs F'\bs F/T)^{1/2}\|\|(\bs F'\bs F)^{-1/2}\bs F' \bs U_{-r} \|\\
        &=O_{p}\left(\frac{1}{N^{\alpha}}\right)+O_{p}\left(\frac{N}{N^{\alpha}}\frac{1}{T}\right)+ O_{p}\left(\frac{1}{\sqrt{N^{\alpha} T}}\right)\\
        &\quad +O_{p}\left(k\delta_{NT}^{-1}\frac{1}{\sqrt{T}}\left[\frac{1}{(N^{\alpha}T)^{1/4}} + \frac{1}{N^{\alpha/2}} + \sqrt{\frac{N}{N^{\alpha}}\frac{1}{T}}\right]\right),
    \end{align*}
    with the first three terms following from \cref{res:VNTr}, and the final terms due to \cref{res:helpF} and the fact that $\frac{1}{NT}\|\bs F'\bs e\|^2 = O_{p}(1)$ by Lemma A.1 in \citet{bai2023approximate}.

    \paragraph{Term $(III)$: \cref{thm:1}, correlation in $\bs e_t$.}
    \begin{equation}
        \begin{split}
            \|\bs e_{T}'\bs M_{2}\bs F\bs\gamma\| & = \frac{1}{\sqrt{NT}}\|\bs e_{T}'\bs V \bs D_{2}\bs U '\bs F\bs\gamma\|\\
            &=\frac{1}{\sqrt{N}}O_{p}\left(\sqrt{\frac{N}{N^{\alpha}}}\right)\|\bs e_{T}'\bs V_{r} \|\|\frac{1}{\sqrt{T}}\bs U_{r}'\bs F\bs\gamma\| \\
            &\quad + \frac{1}{\sqrt{N}}O_{p}(k\delta_{NT}^{-1})\|\bs e_{T}'\bs V_{-r} \|\|\frac{1}{\sqrt{T}}\bs U_{-r} '\bs F\bs\gamma\|\\
            &= 
            O_{p}\left(\frac{1}{N^{\alpha/2}}\right)\\
            &\quad +
            O_{p}\left(\frac{N^{3/2}}{N^{3\alpha/2}}\frac{1}{T}\right)+ O_{p}\left(\frac{N^{1/2}}{N^{3\alpha/2}}\right)+O_{p}\left(\frac{N^2}{N^{2\alpha}T^{1/2}}\right)+O_{p}\left(\frac{N^2}{N^{2\alpha}T^{3/2}}\right)\\
            &\quad + O_{p}\left(k\delta_{NT}^{-1}\left[\frac{1}{(N^{\alpha}T)^{1/4}} + \frac{1}{N^{\alpha/2}} + \sqrt{\frac{N}{N^{\alpha}}\frac{1}{T}}\right]\right),
        \end{split}
    \end{equation}
    where we used \cref{res:etVNTr} to obtain the terms on the second to last line, and \cref{res:helpF} and the fact that $N^{-1/2}\|\bs e_{T}'\bs V_{-r} \| \leq N^{-1/2}\|\bs e_{T}\| = O_{p}(1)$ by \cref{assA1} to obtain the final line.

    \paragraph{Term $(III)$: \cref{thm:3},  serial correlation in $\bs e_t$.}
    If $\bs e_{t}$ is independent with $\{\bs F_{s},\bs e_{s}\}_{s<t}$ then we can derive a tighter bound. In this case, term $(III)$ is mean zero. We then bound
    \begin{equation}
        \begin{split}
            \text{Var}\left(\frac{1}{\sqrt{NT}}\bs e_{T}'\bs V_{r} \bs D_{2,r}\bs U_{r}'\bs F\bs\gamma\right)&\leq C\frac{\lambda_{\max}(\mathbb{E}[\bs e_{T}'\bs e_{T}])}{N^{\alpha}},\\
            \text{Var}\left(\frac{1}{\sqrt{NT}}\bs e_{T}'\bs V_{-r} \bs D_{2,-r}\bs U_{-r} '\bs F\bs\gamma\right)&\leq \frac{1}{N}k^2\delta_{NT}^{-2}\left(O_{p}\left(\frac{1}{\sqrt{N^{\alpha}T}}\right) + O_{p}\left(\frac{1}{N^{\alpha}}\right)+ O_{p}\left(\frac{N}{N^{\alpha}}\frac{1}{T}\right)\right).
        \end{split}
    \end{equation}
    We can then improve the bound on $(III)$ as follows.
    \begin{equation}
        \begin{split}
            \|\bs e_{T}'\bs M_{2}\bs F\bs\gamma\| & \leq  \frac{1}{\sqrt{NT}}\|\bs e_{T}'\bs V_{r} \bs D_{2,r}\bs U_{r}'\bs F\bs\gamma\|+ \frac{1}{\sqrt{NT}}\|\bs e_{T}'\bs V_{-r} \bs D_{2,-r}\bs U_{-r} '\bs F\bs\gamma\|\\
            &= 
            O_{p}\left(\frac{1}{N^{\alpha/2}}\right)+ O_{p}\left(k\delta_{NT}^{-1}\frac{1}{\sqrt{N}}\left[\frac{1}{(N^{\alpha}T)^{1/4}} + \frac{1}{N^{\alpha/2}} + \sqrt{\frac{N}{N^{\alpha}}\frac{1}{T}}\right]\right).
        \end{split}
    \end{equation}

    \paragraph{Term $(IV)$}
    By \cref{ass:varepsilon}, we have ${\bs b}_{T}'\bs M\bs\varepsilon/\sqrt{{\bs b}_{T}'\bs M^2 {\bs b}_{T}} = O_{p}(1)$. Then,
    \begin{equation}
        \begin{split}
            \|(IV)\|&= \sqrt{\bs b_{T}'\bs M^2\bs b_{T}}\frac{\bs b_{T}'\bs M\bs\varepsilon}{\sqrt{\bs b_{T}'\bs M^2\bs b_{T}}}\\
            &=O_{p}\left(\frac{1}{\sqrt{T}}\right)+ O_p\left(k\delta_{NT}^{-2}\right).
        \end{split}
    \end{equation}
    Combining the established bounds yields \cref{thm:1} and \cref{thm:3}.

\newpage
\section{FRED-MD and FRED-QD subset \label{app:sec:innersubset}}

\Cref{app:tbl:innersubset} presents the list of variables that are both in the FRED-MD and the FRED-QD, with the id as listed in the FRED-MD and the FRED-QD.
The column group corresponds to the groups of the FRED-MD: (1) Output and income; (2) Labor market; (3) Housing; (4) Consumption, orders, and inventories; (5) Money and credit; (6) Interest and exchange rates; (7) Prices; (8) Stock market.

The column tcode is the transformation applied to the raw data $x$, as specified in the FRED-QD: (1) no transformation; (2) $\Delta x_t$; (3) $\Delta^2 x_t$; (4) $\log(x_t)$; (5) $\Delta\log(x_t)$; (6) $\Delta^2\log(x_t)$. The transformation code differs between the FRED-MD and FRED-QD for 15 variables. For example, Housing Starts series are suggested to be transformed by taking logs in the FRED-MD, but the recommendation in the FRED-QD is to apply log differences. In those cases, we follow the FRED-QD.

\begin{table}
    \caption{Variables in FRED-MD and FRED-QD \label{app:tbl:innersubset}}
    \resizebox{\textwidth}{!}{
    \begin{tabular}{rrrrll}
        \toprule
            id (M)	&	id (Q)	&	Group (M)	&	tcode (Q)	&	Mnemonic (M)	&	Description	\\
        \midrule
            6	&	22	&	1	&	5	&	INDPRO	&	IP Index	\\
            8	&	23	&	1	&	5	&	IPFINAL	&	IP: Final Products and Nonindustrial Supplies	\\
            9	&	24	&	1	&	5	&	IPCONGD	&	IP: Final Products (Market Group)	\\
            10	&	28	&	1	&	5	&	IPDCONGD	&	IP: Consumer Goods	\\
            11	&	30	&	1	&	5	&	IPNCONGD	&	IP: Durable Consumer Goods	\\
            12	&	31	&	1	&	5	&	IPBUSEQ	&	IP: Nondurable Consumer Goods	\\
            13	&	25	&	1	&	5	&	IPMAT	&	IP: Materials	\\
            14	&	26	&	1	&	5	&	IPDMAT	&	IP: Durable Materials	\\
            15	&	27	&	1	&	5	&	IPNMAT	&	IP: Nondurable Materials	\\
            16	&	194	&	1	&	5	&	IPMANSICS	&	IP: Manufacturing (SIC)	\\
            17	&	195	&	1	&	5	&	IPB51222S	&	IP: Residential Utilities	\\
            18	&	196	&	1	&	5	&	IPFUELS	&	IP: Fuels	\\
            19	&	34	&	1	&	1	&	CUMFNS	&	Capacity Utilization: Manufacturing	\\
            \hline
            20	&	80	&	2	&	1	&	HWI	&	Help-Wanted Index for United States	\\
            21	&	220	&	2	&	2	&	HWIURATIO	&	Ratio of Help Wanted/No. Unemployed	\\
            23	&	57	&	2	&	5	&	CE16OV	&	Civilian Employment	\\
            24	&	59	&	2	&	2	&	UNRATE	&	Civilian Unemployment Rate	\\
            25	&	197	&	2	&	2	&	UEMPMEAN	&	Average Duration of Unemployment (Weeks)	\\
            26	&	65	&	2	&	5	&	UEMPLT5	&	Civilians Unemployed - Less Than 5 Weeks	\\
            27	&	66	&	2	&	5	&	UEMP5TO14	&	Civilians Unemployed for 5-14 Weeks	\\
            29	&	67	&	2	&	5	&	UEMP15T26	&	Civilians Unemployed for 15-26 Weeks	\\
            30	&	68	&	2	&	5	&	UEMP27OV	&	Civilians Unemployed for 27 Weeks and Over	\\
            31	&	221	&	2	&	5	&	CLAIMSx	&	Initial Claims	\\
            32	&	35	&	2	&	5	&	PAYEMS	&	All Employees: Total nonfarm	\\
            33	&	39	&	2	&	5	&	USGOOD	&	All Employees: Goods-Producing Industries	\\
            35	&	42	&	2	&	5	&	USCONS	&	All Employees: Construction	\\
            36	&	37	&	2	&	5	&	MANEMP	&	All Employees: Manufacturing	\\
            37	&	40	&	2	&	5	&	DMANEMP	&	All Employees: Durable goods	\\
            38	&	41	&	2	&	5	&	NDMANEMP	&	All Employees: Nondurable goods	\\
            39	&	38	&	2	&	5	&	SRVPRD	&	All Employees: Service-Providing Industries	\\
            40	&	50	&	2	&	5	&	USTPU	&	All Employees: Trade, Transportation \& Utilities	\\
            41	&	53	&	2	&	5	&	USWTRADE	&	All Employees: Wholesale Trade	\\
            42	&	52	&	2	&	5	&	USTRADE	&	All Employees: Retail Trade	\\
            43	&	44	&	2	&	5	&	USFIRE	&	All Employees: Financial Activities	\\
            44	&	51	&	2	&	5	&	USGOVT	&	All Employees: Government	\\
            45	&	198	&	2	&	2	&	CES0600000007	&	Avg Weekly Hours : Goods-Producing	\\
            46	&	79	&	2	&	2	&	AWOTMAN	&	Avg Weekly Overtime Hours : Manufacturing	\\
            47	&	77	&	2	&	1	&	AWHMAN	&	Avg Weekly Hours : Manufacturing	\\
            120	&	216	&	2	&	6	&	CES0600000008	&	Avg Hourly Earnings : Goods-Producing	\\
            \hline
            48	&	81	&	3	&	5	&	HOUST	&	Housing Starts: Total New Privately Owned	\\
            49	&	85	&	3	&	5	&	HOUSTNE	&	Housing Starts, Northeast	\\
            50	&	84	&	3	&	5	&	HOUSTMW	&	Housing Starts, Midwest	\\
            51	&	86	&	3	&	5	&	HOUSTS	&	Housing Starts, South	\\
            52	&	87	&	3	&	5	&	HOUSTW	&	Housing Starts, West	\\
            53	&	83	&	3	&	5	&	PERMIT	&	New Private Housing Permits (SAAR)	\\
            54	&	227	&	3	&	5	&	PERMITNE	&	New Private Housing Permits, Northeast (SAAR)	\\
            55	&	228	&	3	&	5	&	PERMITMW	&	New Private Housing Permits, Midwest (SAAR)	\\
            56	&	229	&	3	&	5	&	PERMITS	&	New Private Housing Permits, South (SAAR)	\\
            57	&	230	&	3	&	5	&	PERMITW	&	New Private Housing Permits, West (SAAR)	\\
            \hline
            59	&	90	&	4	&	5	&	AMDMNOx	&	New Orders for Durable Goods	\\
            60	&	93	&	4	&	5	&	ANDENOx	&	New Orders for Nondefense Capital Goods	\\
        \bottomrule
    \end{tabular}
    }
\end{table}

\begin{table}\ContinuedFloat
    \caption{Variables in FRED-MD and FRED-QD (continued)}
    \resizebox{\textwidth}{!}{
    \begin{tabular}{rrrrll}
    \toprule
        id (M)	&	id (Q)	&	Group (M)	&	tcode (Q)	&	Mnemonic (M)	&	Description	\\
    \midrule
        61	&	92	&	4	&	5	&	AMDMUOx	&	Unfilled Orders for Durable Goods	\\
        62	&	222	&	4	&	5	&	BUSINVx	&	Total Business Inventories	\\
        63	&	223	&	4	&	2	&	ISRATIOx	&	Total Business: Inventories to Sales Ratio	\\
        123	&	188	&	4	&	1	&	UMCSENTx	&	Consumer Sentiment Index	\\
        \hline
        66	&	161	&	5	&	5	&	M2REAL	&	Real M2 Money Stock	\\
        68	&	199	&	5	&	6	&	TOTRESNS	&	Total Reserves of Depository Institutions	\\
        69	&	200	&	5	&	7	&	NONBORRES	&	Reserves Of Depository Institutions	\\
        73	&	224	&	5	&	2	&	CONSPI	&	Nonrevolving consumer credit to Personal Income	\\
        124	&	217	&	5	&	6	&	DTCOLNVHFNM	&	Consumer Motor Vehicle Loans Outstanding	\\
        125	&	218	&	5	&	6	&	DTCTHFNM	&	Total Consumer Loans and Leases Outstanding	\\
        126	&	219	&	5	&	6	&	INVEST	&	Securities in Bank Credit at All Commercial Banks	\\
        \hline
        78	&	144	&	6	&	2	&	FEDFUNDS	&	Effective Federal Funds Rate	\\
        79	&	225	&	6	&	2	&	CP3Mx	&	3-Month AA Financial Commercial Paper Rate	\\
        80	&	145	&	6	&	2	&	TB3MS	&	3-Month Treasury Bill	\\
        81	&	146	&	6	&	2	&	TB6MS	&	6-Month Treasury Bill	\\
        82	&	147	&	6	&	2	&	GS1	&	1-Year Treasury Rate	\\
        83	&	201	&	6	&	2	&	GS5	&	5-Year Treasury Rate	\\
        84	&	148	&	6	&	2	&	GS10	&	10-Year Treasury Rate	\\
        85	&	150	&	6	&	2	&	AAA	&	Moody's Seasoned Aaa Corporate Bond Yield	\\
        86	&	151	&	6	&	2	&	BAA	&	Moody's Seasoned Baa Corporate Bond Yield	\\
        87	&	226	&	6	&	1	&	COMPAPFFx	&	3-Month Commercial Paper Minus FEDFUNDS	\\
        88	&	202	&	6	&	1	&	TB3SMFFM	&	3-Month Treasury C Minus FEDFUNDS	\\
        91	&	203	&	6	&	1	&	T5YFFM	&	5-Year Treasury C Minus FEDFUNDS	\\
        93	&	204	&	6	&	1	&	AAAFFM	&	Moody's Aaa Corporate Bond Minus FEDFUNDS	\\
        95	&	182	&	6	&	5	&	TWEXAFEGSMTHx	&	Trade Weighted U.S. Dollar Index	\\
        96	&	184	&	6	&	5	&	EXSZUSx	&	Switzerland / U.S. Foreign Exchange Rate	\\
        97	&	185	&	6	&	5	&	EXJPUSx	&	Japan / U.S. Foreign Exchange Rate	\\
        98	&	186	&	6	&	5	&	EXUSUKx	&	U.S. / U.K. Foreign Exchange Rate	\\
        99	&	187	&	6	&	5	&	EXCAUSx	&	Canada / U.S. Foreign Exchange Rate	\\
        \hline
        100	&	122	&	7	&	6	&	WPSFD49207	&	PPI: Finished Goods	\\
        101	&	124	&	7	&	6	&	WPSFD49502	&	PPI: Finished Consumer Goods	\\
        102	&	127	&	7	&	6	&	WPSID61	&	PPI: Intermediate Materials	\\
        103	&	205	&	7	&	6	&	WPSID62	&	PPI: Crude Materials	\\
        104	&	130	&	7	&	5	&	OILPRICEx	&	Crude Oil, spliced WTI and Cushing	\\
        105	&	206	&	7	&	6	&	PPICMM	&	PPI: Metals and metal products	\\
        106	&	120	&	7	&	6	&	CPIAUCSL	&	CPI : All Items	\\
        107	&	207	&	7	&	6	&	CPIAPPSL	&	CPI : Apparel	\\
        108	&	208	&	7	&	6	&	CPITRNSL	&	CPI : Transportation	\\
        109	&	209	&	7	&	6	&	CPIMEDSL	&	CPI : Medical Care	\\
        110	&	210	&	7	&	6	&	CUSR0000SAC	&	CPI : Commodities	\\
        111	&	211	&	7	&	6	&	CUSR0000SAD	&	CPI : Durables	\\
        112	&	212	&	7	&	6	&	CUSR0000SAS	&	CPI : Services	\\
        113	&	213	&	7	&	6	&	CPIULFSL	&	CPI : All Items Less Food	\\
        114	&	214	&	7	&	6	&	CUSR0000SA0L2	&	CPI : All items less shelter	\\
        115	&	215	&	7	&	6	&	CUSR0000SA0L5	&	CPI : All items less medical care	\\
        116	&	95	&	7	&	6	&	PCEPI	&	Personal Cons. Expend.: Chain Index	\\
        \hline
        74	&	246	&	8	&	5	&	S\&P 500	&	S\&P's Common Stock Price Index: Composite	\\
        75	&	245	&	8	&	5	&	S\&P: indust	&	S\&P's Common Stock Price Index: Industrials	\\
        76	&	247	&	8	&	2	&	S\&P div yield	&	S\&P's Composite Common Stock: Dividend Yield	\\
        77	&	248	&	8	&	5	&	S\&P PE ratio	&	S\&P's Composite Common Stock: Price-Earnings Ratio	\\
        127	&	178	&	8	&	1	&	VIXCLSx	&	VIX	\\
    \bottomrule
    \end{tabular}
    }
\end{table}

\end{appendices}

}{}

\end{document}